\documentclass[reprint,nofootinbib,superscriptaddress,amsmath,amssymb,aps,prd]{revtex4-1}
\usepackage{amsmath}
\usepackage{graphicx}
\usepackage{color}
\usepackage{dcolumn}
\usepackage{bm}
\usepackage{slashed}
\usepackage{epstopdf}
\usepackage{multirow}
\usepackage{float}
\usepackage{enumitem}
\usepackage[colorlinks,linkcolor=red,anchorcolor=green,citecolor=blue,CJKbookmarks=True]{hyperref}
\RequirePackage{mathptmx}
\RequirePackage{amssymb,amsfonts}
\usepackage{lipsum} % for dummy text
\DeclareSymbolFont{symbols} {OMS}{cmsy}{m}{n}
\newcommand{\bea}{\begin{eqnarray*}}
\newcommand{\eea}{\end{eqnarray*}}
\newcommand{\bean}{\begin{eqnarray}}
\newcommand{\eean}{\end{eqnarray}}

\newcommand{\meq}[1]{(\ref{#1})}

\newcommand{\non}{\nonumber \\}

\newcommand{\hsp}{\hspace{0.1mm}}

\newcommand{\pp}{\partial}

\begin{document}
\title{Kerr-MOG-(A)dS black hole and its shadow in scalar-tensor-vector gravity theory}

\author{Wentao Liu}
\affiliation{Department of Physics, Key Laboratory of Low Dimensional Quantum Structures and Quantum Control of Ministry of Education, and Synergetic Innovation Center for Quantum Effects and Applications, Hunan Normal
University, Changsha, Hunan 410081, P. R. China}

\author{Di Wu}
%\email[]{wdcwnu@163.com}
\affiliation{School of Physics and Astronomy, China West Normal University, Nanchong, Sichuan 637002, P. R. China}

\author{Xiongjun Fang}
\email[]{fangxj@hunnu.edu.cn} \affiliation{Department of Physics, Key Laboratory of Low Dimensional Quantum Structures and Quantum Control of Ministry of Education, and Synergetic Innovation Center for Quantum Effects and Applications, Hunan Normal
University, Changsha, Hunan 410081, P. R. China}
%\affiliation{GCAP-CASPER, Department of Physics, Baylor University, One Bear Place \#97316, Waco, Texas 76798-7316, USA}

\author{Jiliang Jing}
%\email[]{jljing@hunnu.edu.cn}
\affiliation{Department of Physics, Key Laboratory of Low Dimensional Quantum Structures and
Quantum Control of Ministry of Education, and Synergetic Innovation Center for Quantum Effects and Applications, Hunan Normal University, Changsha, Hunan 410081, P. R. China}

\author{Jieci Wang}
\email[]{jcwang@hunnu.edu.cn} \affiliation{Department of Physics, Key Laboratory of Low Dimensional Quantum Structures and Quantum Control of Ministry of Education, and Synergetic Innovation Center for Quantum Effects and Applications, Hunan Normal
University, Changsha, Hunan 410081, P. R. China}

\begin{abstract}
The scalar-tensor-vector gravity (STVG) theory has attracted significant interest due to its ability to effectively address the issue of galaxy rotation curves and clusters of galaxies without considering the influence of dark matter. In this paper, we construct rotating black hole solutions with a cosmological constant in the STVG theory (i.e., Kerr-MOG-(A)dS black hole solutions), where the import of a gravitational charge as a source modifies the gravitational constant, determined by $ G=G_{\text{N}}(1+\alpha) $. For Kerr-MOG-dS spacetime, the observer is situated at a specific location within the domain of outer communication, rather than being located infinitely far away. Since black hole shadows are shaped by light propagation in spacetime, the interaction between the MOG parameter and the cosmological constant is expected to produce novel effects on these shadows. 
As the cosmological constant $\Lambda$ increases, the apparent size of the black hole shadow decreases.
Additionally, the shadow expands with an increase in the MOG parameter $\alpha$, reaching a maximum at a certain value, and its shape becomes more rounded under an arbitrary rotation parameter, which leads to degeneracy between different black hole parameters. However, by employing numerical ray-tracing techniques, we have found that gravitational lensing and the frame-dragging effect effectively distinguish this degeneracy. Our work contributes to a deeper understanding of black holes in modified gravity, their observational signatures, and constraints.
\end{abstract}

\maketitle

%%%%%%%%%%%%%%%%%%%%%%
\section{Introduction}
%%%%%%%%%%%%%%%%%%%%%%

Although the no-hair theorem suggests that black holes are characterized by their mass, spin, and electric charge, however, the electric charge is usually negligible for astrophysical black holes. The reason is that a black hole with mass $ M $ and electric charge $ Q $ will not gravitationally attract particles with mass $ m $ and electric charge $ e $ if $eQ > Mm$. Given the $ m/e $ ratio for electrons is approximately $ 10^{-21} $, it is highly unlikely for black holes to maintain a significant electric charge \cite{Gibbons:1975kk}. Moreover, processes including vacuum polarization, pair production, and neutralization ensure that stellar-mass black holes cannot sustain significant electric charge. Any accumulated charge dissipates quickly and escapes observation by gravitational wave detectors \cite{Gibbons:1975kk,Blandford:1977ds,Cardoso:2016olt}. However, by performing general-relativistic simulations of charged black holes, Bozzola et al. have proposed that the gravitational wave event GW150914 could have a charge-to-mass ratio as high as $0.3$ \cite{Bozzola:2020mjx,Gupta:2021rod,Carullo:2021oxn}. Consequently, the term ``charge'' could be more appropriately interpreted as dark charge or gravitational charge, sparking renewed interest in the scalar-tensor-vector gravity (STVG) theory, also known as modified gravity (MOG) theory, which was first proposed nearly two decades ago \cite{Moffat:2005si}. The STVG theory introduces additional massive vector and scalar fields to the metric tensor field, positing that the associated ``charge'' is a gravitational charge denoted by $Q_5\equiv\kappa M$, which is determined by a body's mass $M$ and its interaction with the vector field.

On the observational side, despite numerous attempts to detect dark matter, all experimental efforts up to now have yet to achieve recognized success \cite{An:2022hhb,PandaX:2023toi}. Therefore, on the theoretical side, STVG theory has attracted a great deal of attention because it not only successfully explains the rotation curves of galaxies and the properties of galaxy clusters, but also does so without considering dark matter \cite{Moffat:2013sja,Moffat:2013uaa, Moffat:2014pia,Moffat:2014bfa, Rahvar:2022yhj}. In addition, observational data from the Event Horizon Telescope (EHT) on the shadows of Sgr A* and M87* \cite{EventHorizonTelescope:2019dse,EventHorizonTelescope:2019ggy,EventHorizonTelescope:2022xqj} support the predictions of STVG theory and there are criteria to distinguish this theory from general relativity \cite{Moffat:2019uxp}. The axially symmetric solution in STVG theory provided by Moffat et al. is called the Kerr-MOG black hole \cite{Moffat:2014aja}. To date, several new black hole solutions related to this theory have been obtained \cite{Moffat:2018jmi,Moffat:2021tfs,Liu:2023MOG}. Building on the established solutions in STVG theory, extensive research has been conducted on various aspects of black holes, including their thermodynamic properties \cite{Mureika:2015sda}, geodesic motion \cite{Lee:2017fbq}, shadows \cite{Moffat:2015kva,Wang:2018prk,Wei:2018aft,Guo:2018kis, Sheoran:2017dwb,Zhang:2024jrw,Qiao:2020fta,Konoplya:2021slg,Younsi:2016azx}, and quasi-normal modes \cite{Manfredi:2017xcv,Jiang:2024sze,Liu:2022dcn}, etc.

Despite much progress in the last few years in constructing black hole solutions in STVG theory and investigating their physical properties, rotating (A)dS black holes in STVG theory remain to be the virgin territory and thus need to be explored deeply. On the other hand, our current universe is in a state of accelerated expansion, and the simplest explanation for such accelerated expansion is a positive cosmological constant \cite{SupernovaSearchTeam:1998fmf,
SupernovaCosmologyProject:1998vns}. Thus, it is significant to construct rotating dS black hole solutions in STVG theory. Additionally, the study of rotating AdS black holes has already shed light on the nature of gravity through gauge-gravity dualities \cite{Maldacena:1997re,Gubser:1998bc,Witten:1998qj}, so it is very important and remarkable to seek the rotating AdS black hole solution in STVG theory. These three aspects motivate us to conduct the present work.

In this work, we aim to extend the Kerr-MOG solution \cite{Moffat:2014aja} to more general cases with a cosmological constant, thereby obtaining Kerr-MOG-(A)dS black holes. Given the recent groundbreaking observations by the EHT, theoretical research on the properties of black holes frequently focuses on the analysis of black hole images \cite{Chen:2022scf,Liu:2024lve}. Hence, we have also investigated the effects of the MOG parameter and a positive cosmological constant on black hole shadows, as well as the numerical estimation of the angular radius of the supermassive black holes Sgr A* and M87* within this theoretical framework. The remaining part of this paper is organized as follows. In Sec. \ref{Sec.2}, we provide a concise overview of the STVG theory and give the Kerr-like solution of the field equations with a cosmological constant, i.e., Kerr-MOG-(A)dS black holes. In Sec. \ref{Sec.3}, utilizing the geodesic equation, we derive the orbital equations for photons in the context of Kerr-MOG-dS spacetime. In Sec. \ref{Sec.4}, we examine the shadows of black holes, with a primary focus on the apparent shapes and distortions of the shadow. Also, we numerically calculate these two observables, $ R_s $ and $ \delta_s $, to characterize the shadow. Sec. \ref{Sec.5} focuses on providing an overview of our conclusions while also contemplating the possible development and exploration of related research topics in the future.

%%%%%%%%%%%%%%%%%%%%%%%%%%%%%%%%%%%%%%%%%%%%%%%%%%%%%%%%%%
\section{STVG theory and black hole solution}\label{Sec.2}
%%%%%%%%%%%%%%%%%%%%%%%%%%%%%%%%%%%%%%%%%%%%%%%%%%%%%%%%%%

In this section, we first briefly review the STVG theory and then proceed to construct new rotating black hole solutions using the Kerr-Schild-(A)dS metric ansatz as a seed solution.

%%%%%%%%%%%%%%%%%%%%%%%%
\subsection{STVG theory}
%%%%%%%%%%%%%%%%%%%%%%%%%

The action corresponding to STVG theory is given by \cite{Moffat:2005si}
\begin{align}
\mathcal{S}=\mathcal{S}_\text{M}+\mathcal{S}_\text{G}+\mathcal{S}_\phi+\mathcal{S}_\text{S},
\end{align}
with
\begin{equation}
\begin{aligned}
\mathcal{S}_\text{G}=&\frac{1}{16\pi}\int d^4x \sqrt{-g}\left[\frac{1}{G}\left(R-2\Lambda\right)\right],\\
\mathcal{S}_\phi=&\int d^4x \sqrt{-g}\left(-\frac{1}{4}B^{ab}B_{ab}+\frac{1}{2}\mu^2\phi^a\phi_a\right),\\
\mathcal{S}_\text{S}=&\int d^4x\sqrt{-g}\left[\frac{1}{G^3}\left(\frac{1}{2}g^{ab}\nabla_a G\nabla_b G -V(G)\right)\right.\\
&\left.+\frac{1}{\mu^2 G}\left(\frac{1}{2}g^{ab}\nabla_a\mu\nabla_b\mu-V(\mu)\right)\right],
\end{aligned}
\end{equation}
where the symbol $ \mathcal{S}_\text{M} $ represents the matter action, $ \mathcal{S}_\text{G} $ represents the action for Einstein gravity, and $ R $ and $ \Lambda $ are the Ricci scalar and cosmological constant, respectively. Additionally, the action $ \mathcal{S}_\phi $ describes a Proca field $ \phi^a $ with mass $ \mu $. The field strength, represented by $ B_{ab}=\pp_a\phi_b-\pp_b\phi_a $, satisfies the following equations:
\begin{align}\label{motionEq}
\nabla_bB^{ab}=\frac{1}{\sqrt{-g}}\pp_b\left(\sqrt{-g}B^{ab}\right)=0,\\
\pp_cB_{ab}+\pp_aB_{bc}+\pp_bB_{ca}=0.
\end{align}
The potentials $V(G)$ and $V(\mu)$ are linked to the scalar fields $G(x)$ and $\mu(x)$ within the action $ \mathcal{S}_\mathrm{S} $, respectively.

Under the assumption of vacuum solution, the action $ \mathcal{S} $ can be simplified to:
\begin{equation}
\begin{aligned}\label{action}
\mathcal{S}=&\int d^4x\sqrt{-g}\Big[\frac{1}{16\pi G}\left(R-2\Lambda\right)\\
&-\frac{1}{4}B^{ab}B_{ab}+\frac{1}{2}\mu^2\phi^a\phi_a \Big].
\end{aligned}
\end{equation}
The field equation of motion derived from varying the action in Eq. \meq{action} with respect to the metric is given by
\begin{align}\label{gravityEq}
\mathcal{G}_{ab}+\Lambda g_{ab}=-8\pi G T_{ab},
\end{align}
where $ \mathcal{G}_{ab}= R_{ab}-\frac{1}{2}Rg_{ab}$ is Einstein tensor, and the energy-momentum tensor is
\begin{align}
\begin{aligned}
T_{ab}=&-\frac{1}{4\pi}\Big(B_a\hsp^cB_{bc}-\frac{1}{4}g_{ab}B^{cd}B_{cd}\Big)\\
&+\frac{\mu^2}{4\pi}\Big( \phi_a\phi_b-\frac{1}{2}g_{ab}\phi^c\phi_c \Big).
\end{aligned}
\end{align}
The first and second terms of the energy-momentum tensor are of the order $\sim(\pp \phi)^2$ and $\sim\mu^2 \phi^2$, respectively. To solve the equation \eqref{gravityEq}, one can typically use the weak field approximation and neglect the higher-order terms including $g^2$, $g\phi$, and $\phi^2$ \cite{Moffat:2013sja,Rahvar:2022yhj}. Concurrently, according to Refs. \cite{Moffat:2013uaa}, the particle mass of the $ \phi^a $ field in the present universe can be fitted as $ m_\phi\sim 10^{-28}eV $, making it negligible for a black hole solution. In STVG theory, it is important to note that the fifth force charge $ Q_5 $ is tied to a particle's inertial mass as $ Q_5=\kappa M $. This parameter $ \kappa $ affects the gravitational constant, given by $ G=G_\text{N}+\kappa^2 $, with the convention $ \kappa^2=\alpha G_\text{N} $. Then, we can obtain the relationship as $ G=(1+\alpha)G_\text{N} $, which can also be thought of as a constant unaffected by the spacetime coordinates. By setting $ \alpha $ to zero, the theory reduces to General Relativity (GR). Thus, $ \alpha $ can be considered as a parameter that measures the deviation of STVG from GR.

In the static case, assuming the vector field $\phi_a$ takes the form \cite{Moffat:2014aja}
\begin{align}\label{phias}
\phi_a= \left(-\sqrt{\alpha G_\text{N}}\frac{M}{r},0,0,0\right),
\end{align}
and setting $G_\text{N}=1$ and $\beta=\alpha/(1+\alpha)$, the corresponding asymptotic (A)dS metric solution is given by \cite{Liu:2023MOG}
\begin{equation}\label{ds2MOG}
\begin{aligned}
ds^2=&-\left(1-\frac{2M_\text{D}}{r}+\frac{\beta M^2_\text{D}}{r^2}-\frac{\Lambda}{3}r^2\right)dt^2\\
&+\left(1-\frac{2M_\text{D}}{r}+\frac{\beta M^2_\text{D}}{r^2}-\frac{\Lambda}{3}r^2\right)^{-1}dr^2\\
&+r^2d\vartheta^2+r^2\sin^2\vartheta d\varphi^2.
\end{aligned}
\end{equation}
The symbol $M_\text{D}$ stands for the ADM mass, which can be related to the Newtonian mass $M$ through the equation $M_\text{D}=(1+\alpha)M$ \cite{Sheoran:2017dwb}.

%%%%%%%%%%%%%%%%%%%%%%%%%%%%%%%%%%%%%
\subsection{Rotating-(A)dS solutions}
%%%%%%%%%%%%%%%%%%%%%%%%%%%%%%%%%%%%%

Rotating black hole solutions play a crucial role in astrophysics. Considering the total spin angular momentum represented by $J=Ma$, adjustments to the vector potential and metric, equations \eqref{phias} to \eqref{ds2MOG}, are necessary to incorporate the spin parameter. In order to obtain the axisymmetric solutions for asymptotically (A)dS spacetimes, we adopt the Kerr-Schild-(A)dS ansatz \cite{Gibbons:2004uw,Malek:2010mh}
\begin{align}
g_{ab}=\bar{g}_{ab}-2\mathcal{H}k_ak_b,
\end{align}
where $ \mathcal{H} $ is an arbitrary scalar function. The vector $ \mathbf{k} $ is null with respect to both the full metric $ g_{ab} $ and the Kerr-(A)dS metric $ \bar{g}_{ab} $, with the latter serving as the background. Alternatively, by introducing the coordinate $(\bar{t}, \bar{r}, \bar{\vartheta}, \bar{\varphi})$, the line elements can be expressed as follows:
\begin{equation}\label{dsKS2}
\begin{aligned}
d\bar{s}^2=&-\left(1-\frac{\Lambda}{3}\bar{r}^2\right)\frac{\Delta_{\bar{\vartheta}}}{\chi}d\bar{t}^2+\frac{\Sigma}{\Delta_{\bar{r}}}d\bar{r}^2\\
&+\frac{\Sigma}{\Delta_{\bar{\vartheta}}}d{\bar{\vartheta}}^2+\frac{(\bar{r}^2+a^2)\sin^2{\bar{\vartheta}}}{\chi}d{\bar{\varphi}}^2\\
&-2\mathcal{H}\left(\frac{\Delta_{\bar{\vartheta}}}{\chi}d\bar{t}+\frac{\Sigma}{\Delta_{\bar{r}}}d\bar{r}-\frac{a\sin^2{\bar{\vartheta}}}{\chi}d{{\bar{\varphi}}}\right)^2,
\end{aligned}
\end{equation}
with
\begin{equation}
\begin{aligned}
\Sigma=&\bar{r}^2+a^2\cos^2{\bar{\vartheta}},& \Delta_{\bar{\vartheta}}=&1+\frac{\Lambda}{3}a^2\cos^2{\bar{\vartheta}},\\
\chi=&1+\frac{\Lambda}{3}a^2,& \Delta_{\bar{r}}=&(\bar{r}^2+a^2)\left(1-\frac{\Lambda}{3}\bar{r}^2\right).
\end{aligned}
\end{equation}
For the metric ansatz \meq{dsKS2}, the Ricci scalar is
\begin{equation}
\begin{aligned}
R=g^{ab}R_{ab}=4\Lambda-2\left(\frac{\pp^2}{\pp \bar{r}^2}+\frac{4\bar{r}}{\Sigma}\frac{\pp}{\pp \bar{r}}
+\frac{2}{\Sigma}\right)\mathcal{H},
\end{aligned}
\end{equation}
and the vacuum gravitational field equations (with cosmological constant) are $ R=4\Lambda $. Therefore, one can obtain the following partial differential equation for the $\mathcal{H}$ function:
\begin{align}
\left(\frac{\pp^2}{\pp \bar{r}^2}+\frac{4\bar{r}}{\Sigma}\frac{\pp}{\pp \bar{r}}+\frac{2}{\Sigma}\right)\mathcal{H}=0.
\end{align}
The above equation has the general solution,
\begin{align}\label{HHH}
\mathcal{H}=\frac{GQ_5^2}{2\Sigma}-\frac{GM\bar{r}}{\Sigma},
\end{align}
where $GM$ and $GQ_5^2$ are constants of integration, as in Ref. \cite{Moffat:2014aja}. In STVG theory, the charge $Q_5>0$ represents a fifth force with a gravitational origin and is proportional to the mass of the source particle, given by $Q_5^2 =\alpha G_\text{N}M^2$. Together with the relation $G=G_\text{N}(1+\alpha)$, $ \mathcal{H} $ can be rewritten as follows:
\begin{align}\label{HHHH}
\mathcal{H}=\frac{G^2_\text{N}M^2\alpha\left(1+\alpha\right)}{2\Sigma}-\frac{G_\text{N}M\bar{r}(1+\alpha)}{\Sigma}.
\end{align}
Then, the seed metric \meq{dsKS2} with scalar function \meq{HHHH} can be expressed as
\begin{align}\label{dsKS3}
ds^2=&-\frac{\Delta_{\bar{\vartheta}}}{\chi}\left(1-\frac{\Lambda \bar{r}^2}{3}+\frac{h\Delta_{\bar{\vartheta}}}{\chi \Sigma}\right)d\bar{t}^2
+\frac{\Sigma}{\Delta_{\bar{r}}}\left(1-\frac{h}{\Delta_{\bar{r}}}\right)d\bar{r}^2\non
&+\frac{\Sigma}{\Delta_{\bar{\vartheta}}}d\bar{\vartheta}^2+\left(\frac{\bar{r}^2+a^2}{\chi}-\frac{ha^2\sin^2\bar{\vartheta}}{\chi^2\Sigma}\right)\sin^2\bar{\vartheta}d\bar{\varphi}^2\non
&-\frac{2h\Delta_{\bar{\vartheta}}}{\chi \Delta_{\bar{r}}}d\bar{t}d\bar{r}
+\left(\frac{\Delta_{\bar{\vartheta}}d\bar{t}}{\chi\Sigma}+\frac{d\bar{r}}{\Delta_{\bar{r}}}\right)\frac{2ha\sin^2\bar{\vartheta}}{\chi}d\bar{\varphi},
\end{align}
where
\begin{align}
h=G_\text{N}M(1+\alpha)(G_\text{N}M\alpha-2\bar{r}).
\end{align}
The non-zero components of the effective tensor, defined as $E_{ab}=R_{ab}-\frac{1}{2}Rg_{ab}+\Lambda g_{ab}+8\pi G T_{ab}$, associated with the metric, are
\begin{equation}
\begin{aligned}\label{E00}
&E_{00}=G_\alpha \Delta_{\bar{\vartheta}}\left[(1-\bar{r}^2\Lambda/3)\Sigma_2+h\Delta_{\bar{\vartheta}} \right]+8\pi G T_{00},\\
%\label{E01}
&E_{01}=G_\alpha h\chi\Delta_{\bar{r}}^{-1}\Delta_{\bar{{\vartheta}}} +8\pi G T_{01},\\
%\label{E03}
&E_{03}=-G_\alpha\Delta_{\bar{\vartheta}}\left(h+2\Delta_{\bar{r}}\right)a\sin^2\bar{\vartheta}+8\pi G T_{03},\\
%\label{E11}
&E_{11}=G_\alpha\left(h-\Delta_{\bar{r}}\right)\chi^2\Delta_{\bar{r}}^{-2}\Sigma^2+8\pi G T_{11},\\
%\label{13}
&E_{13}=-G_\alpha h \chi \Delta_{\bar{\vartheta}}^{-1}\Sigma a\sin^2\bar{\vartheta}+8\pi G T_{13},\\
%\label{22}
&E_{22}=G_\alpha \chi^2\Delta_{\bar{\vartheta}}^{-1}\Sigma^2+8\pi GT_{22},\\
%\label{33}
&E_{33}=G_\alpha\sin^2\bar{\vartheta}\left[(\bar{r}^2+a^2)\Sigma_2+ha^2\sin^2\bar{\vartheta}\right]+8\pi GT_{33},
\end{aligned}
\end{equation}
where
\begin{equation}
\begin{aligned}
G_\alpha&=G_\text{N}^2M^2\alpha(1+\alpha)\chi^{-2}\Sigma^{-3},\\
\Sigma_2&=\Sigma+a^2 \Delta_{\bar{\vartheta}}-a^2\left(1-r^2\Lambda/3\right)(\cos^2\bar{\vartheta}-\sin^2\bar{\vartheta}).
\end{aligned}
\end{equation}

The physical significance of these parameters becomes clearer in the Boyer-Lindquist coordinates $ (t,r,\vartheta,\varphi) $, which offer a more appropriate asymptotic form for the metric. Applying the Boyer-Lindquist transformation,
\begin{equation}
\begin{aligned}
d\bar{r}=&dr, ~~~d{\bar{\vartheta}}=d\vartheta,\\
d\bar{t}=&dt-\frac{h}{\left(1-\frac{\Lambda}{3}r^2\right)\left(\Delta_{\bar{r}}+h\right)}dr,
\\d\bar{\varphi}=&d\varphi -\frac{\Lambda}{3}a dt-\frac{ah}{\left(r^2+a^2\right)\left(\Delta_{\bar{r}}+h\right)}dr,
\end{aligned}
\end{equation}
the metric \meq{dsKS3} can be rewritten in the following form,
\begin{equation}
\begin{aligned}\label{ds2MOG2}
ds^2=&-\frac{\Delta_r}{\Sigma}\left(dt-\frac{a\sin^2\vartheta}{\chi}d\varphi\right)^2
+\frac{\Sigma}{\Delta_r}dr^2+\frac{\Sigma}{\Delta_\vartheta}d\vartheta^2\\
&+\frac{\Delta_\vartheta \sin^2\vartheta}{\Sigma}\left(adt-\frac{(r^2+a^2)}{\chi}d\varphi\right)^2,
\end{aligned}
\end{equation}
with
\begin{align*}
\Delta_\vartheta=\Delta_{\bar{\vartheta}},~~~~~
\Delta_r=\Delta_{\bar{r}}-2G_\text{N} M_\text{D} r+\frac{\alpha G_\text{N}^2 M^2_\text{D}}{(1+\alpha)}.
\end{align*}

To ensure the effective tensor $E_{ab}=0$ is satisfied, we can simply obtain the MOG vector potential as:
\begin{align}\label{MOGfield}
\phi=\phi_adx^a=\frac{M\sqrt{G_\text{N} \alpha }r}{\Sigma}\left(-dt+\frac{a\sin^2\vartheta}{\chi}d\varphi\right).
\end{align}
In Appendix \ref{AppendixA}, we provide the specific forms of all the energy-momentum tensors contained in equations \meq{E00}, which are sourced from the MOG vector potential \meq{MOGfield}. Equations \meq{ds2MOG2} and \meq{MOGfield} describe the Kerr-MOG-(A)dS black hole solutions. When $\Lambda=0$, the solutions recovers the usual Kerr-MOG solution \cite{Moffat:2014aja}; when $a=0$, it reduces to the static (A)dS cases \cite{Liu:2023MOG}. This metric \meq{ds2MOG2} has the similar form as the Kerr-Newman-(A)dS metric. However, in this case, the ADM mass $M_{\text{D}}$ is associated with the coupling parameter $\alpha$, and the source of the vector field is gravitational. This distinction is crucial because astrophysical objects, including black holes, are electrically neutral.

The equations of the horizons are given by:
\begin{align}
\Delta_r=\frac{\Lambda}{3}\left(r-r_m\right)\left(r-r_h\right)\left(r_c-r\right)\left(r+r_b\right)=0,
\end{align}
where $r_m$, $r_h$, and $ r_c $ correspond to the Cauchy horizon, event horizon, and dS cosmological horizon, respectively \cite{Liu:2023MOG,Liu:2024Lv}. The parameters $r_b$, $r_m$, $r_h$, and $r_c$ satisfy the condition $r_b + r_m + r_h + r_c = 0$. Moreover, $r_c \geq r_h > r_m > 0$ and $r_b$ is negative.

%%%%%%%%%%%%%%%%%%%%%%%%%%%%%%%%%%%%
\section{Photon orbits}\label{Sec.3}
%%%%%%%%%%%%%%%%%%%%%%%%%%%%%%%%%%%%

In this section, we provide a brief derivation of photon trajectories in the gravitational field of Kerr-MOG-(A)dS black holes. When a black hole is located between the observer and the extended background source, not all photons emitted by the source can reach the observer after being deflected by the black hole gravitational field. The apparent shape of the black hole is represented by its shadow contour, which is closely related to the geodesics of photons within the black hole spacetime.

The geodesics corresponding to photons geometry are defined by the Hamilton-Jacobi equation:
\begin{align}\label{EqHJ}
\frac{\pp \mathcal{S}}{\pp \tau}=-\frac{1}{2}g^{ab}\frac{\pp \mathcal{S}}{\pp x^a}\frac{\pp \mathcal{S}}{\pp x^b},
\end{align}
where $ \tau $ is an affine parameter along the null geodesics, and $ \mathcal{S} $ is the Jacobi action. We can set up the Killing fields $ \xi^a=(\pp/\pp t)^a $ and $ \psi^a=(\pp/\pp \varphi)^a $, then the corresponding energy $ \mathcal{E} $ and the angular momentum in the direction of the axis of symmetry $ L $ can be defined as \cite{Waldbook,Zeng:2021mok,Liu2023}:
\begin{align}\label{gabE}
\mathcal{E}=&-g_{ab}\xi^a\dot{x}^b=-g_{00}\dot{t}-g_{03}\dot{\varphi},\\ \label{gabL}
L=&g_{ab}\psi^a\dot{x}^b=g_{03}\dot{t}+g_{33}\dot{\varphi}.
\end{align}

When $ \mathcal{S} $ is separable, the Jacobi action for the photons can be expressed in a straightforward and general form:
\begin{align}\label{actionS}
\mathcal{S}=-\mathcal{E}t+L\varphi+\mathcal{S}_r(r)+\mathcal{S}_\vartheta(\vartheta).
\end{align}
Combining Eqs. \meq{EqHJ} and \meq{actionS} and setting $ G_\text{N}=1$ , we obtain two equations of motion for the propagating photons
\begin{align}\label{taut}
\dot{t}=&\frac{\left(r^2+a^2\right)k}{\Sigma \Delta_r}
+\frac{a\left(L\chi-a\mathcal{E}\sin^2\vartheta \right)}{\Sigma \Delta_\vartheta},\\ \label{tauvarphi}
\dot{\varphi}=&\frac{a\chi k}{\Sigma\Delta_r}
+\frac{\chi\left(L\chi\csc^2\vartheta -a\mathcal{E}\right)}{\Sigma \Delta_\vartheta},
\end{align}
with
\begin{align}
k=\left(r^2+a^2\right)\mathcal{E}-aL\chi.
\end{align}
Substituting Eqs. \meq{taut} and \meq{tauvarphi} into the null geodesics equation $ 0=-g_{ab}\dot{x}^a\dot{x}^b $, one can obtain
\begin{align}
\frac{k^2-\rho^4\dot{r}^2}{\Delta_r}-\frac{\rho^4\dot{\vartheta}^2+\left(L\chi\csc\vartheta-a\mathcal{E}\sin\vartheta\right)^2}{\Delta_\vartheta}=0.
\end{align}
By using the Carter constant $ \mathcal{K} $ to separate variables in the above equation, we can obtain two additional equations of motion for photons \cite{Carter:1968rr}
\begin{align}\label{taur}
\dot{r}=&\frac{1}{\Sigma}\sqrt{\mathcal{R}(r)},\\ \label{tauvartheta}
\dot{\vartheta}=&\frac{1}{\Sigma}\sqrt{\Theta(\vartheta)}.
\end{align}
The explicit forms of the radial and angular equations are as follows:
\begin{align}
\mathcal{R}=&k^2-\left[\mathcal{K}+\left(L\chi-a\mathcal{E}\right)^2\right]\Delta_r,\\
\Theta=&\left[\mathcal{K}+\left(L\chi-a\mathcal{E}\right)^2\right]\Delta_\vartheta-\left(a\mathcal{E}\sin\vartheta-L\chi\csc\vartheta\right)^2.
\end{align}
These equations determine the behavior of light propagation within the spacetime influenced by rotating black holes.

The radial motion equation of the photon is of particular interest, as the silhouette of the black hole shadow can be derived from the orbits with constant $r=r_p$ that fulfill the conditions
\begin{align}\label{YGDTJ}
\mathcal{R}(r)\big|_{r=r_p}=0,&&\frac{d}{dr}\mathcal{R}(r)\big|_{r=r_p}=0.
\end{align}
To determine the contour of black hole shadow, one can introduce the impact parameters $\xi=L/\mathcal{E}$ and $\eta=\mathcal{K}/\mathcal{E}^2$, and solve the system of equations accordingly \cite{Chandrasekhar}. The resulting impact parameters will provide the necessary information:
\begin{align}
\xi=&\frac{M_\mathrm{D}\left(a^2+2M\alpha r_p-3r^2_p\right)+\chi r_p\left(a^2+r^2_p\right)}
{a\chi\left[M-r_p+M\alpha+\frac{1}{3}r_p\Lambda\left(a^2+2r_p^2\right)  \right]},
\\
\eta=&\frac{r_p^2\mathcal{W}-M_\mathrm{D}\Lambda r_p^5\left(2a^2+\frac{4}{3}M\alpha r_p\right)+r_p^6-r_p^6\chi^2}
{a^2\left[M-r_p+M\alpha+\frac{1}{3}r_p\Lambda\left(a^2+2r_p^2\right)  \right]^2},
\end{align}
where
\begin{equation}
\begin{aligned}
\mathcal{W}=&4M r_p(1+\alpha)\left[a^2 +M^2 \alpha (1+\alpha)\right]\\
&-4M^2\alpha(1+\alpha)\Delta_r-r_p^2\left[r_p-3M(1+\alpha)\right]^2.
\end{aligned}
\end{equation}
These two impact parameters are equivalent to the results provided in Ref. \cite{Wang:2018prk} when $\Lambda=0$. For simplicity, we set $M=1$ in subsequent calculations.

%%%%%%%%%%%%%%%%%%%%%%%%%%%%%%%%%%%%%%%%%%%%%
\section{SHADOWS OF BLACK HOLES}\label{Sec.4}
%%%%%%%%%%%%%%%%%%%%%%%%%%%%%%%%%%%%%%%%%%%%%

While AdS solutions play a crucial role in AdS/CFT correspondences, they lack empirical support from specific astronomical observations. Therefore, in this section, we primarily focus on the black hole shadows in Kerr-MOG-dS spacetime. For this, we need to adopt the analysis approach of A. Grenzebach \cite{Grenzebach:2014fha} and the numerical ray-tracing method \cite{Wang:2017qhh,Wang:2021ara,Hu:2020usx,Zhong:2021mty,Chen:2023wna} to investigate the shadow of Kerr-MOG black holes in the presence of a positive cosmological constant.

\subsection{Apparent shape}\label{4.1}

In the Kerr-MOG-dS spacetime, there exist two unstable light orbits within the domain of outer communication, which extends from the outer horizon $r_h$ to the cosmological horizon $r_c$: a prograde orbit and a retrograde orbit. Both orbits serve as limit curves for lightlike geodesics that spiral towards them. Similarly to the circular lightlike geodesics in the Schwarzschild spacetime \cite{Perlick:2021aok}, all spherical lightlike geodesics in the domain of outer communication are also unstable and can serve as limit curves for past-oriented light rays from an observer position that spiral towards them. Hence, the photon region determines the boundary curve of the black hole shadow. Assume that the observer is positioned at $ (r_0, \vartheta_0) $, where  $r_h < r_0 < r_c$. We define an orthonormal tetrad as \cite{Eiroa:2017uuq}:
\begin{equation}\label{zjbj}
\begin{aligned}
e_0=&\left.\frac{r^2+a^2}{\sqrt{\Sigma\Delta_r}}\pp_t+\frac{a\chi}{\sqrt{\Sigma\Delta_r}}\pp_\varphi\right|_{(r_0,\vartheta_0)},\\
e_1=&\left.\sqrt{\frac{\Delta_\vartheta}{\Sigma}}\pp_\vartheta\right|_{(r_0,\vartheta_0)},\\
e_2=&\left.-\frac{\chi \csc\vartheta}{\sqrt{\Sigma\Delta_\vartheta}}\pp_\varphi-\frac{a\sin\vartheta}{\sqrt{\Sigma\Delta_\vartheta}}\pp_t\right|_{(r_0,\vartheta_0)},\\
e_3=&\left.-\sqrt{\frac{\Delta_r}{\Sigma}}\pp_r\right|_{(r_0,\vartheta_0)}.
\end{aligned}
\end{equation}
It is easy to verify that $e_0$, $e_1$, $e_2$, and $e_3$ are orthonormal. In this chosen tetrad, $e_0$ represents the observer four-velocity,  $e_3$ corresponds to the spatial direction towards the center of the black hole, and the combination $e_0\pm e_3$ is tangential to the principal null congruences of the metric.
For each light ray $ \lambda(s) $ with coordinate representation $ t(s), r(s),\vartheta(s),\varphi(s) $, the general form of tangent vector is
\begin{align}\label{lambda1}
\dot{\lambda}=\dot{t}\pp_t+\dot{r}\pp_r+\dot{\vartheta}\pp_\vartheta+\dot{\varphi}\pp_\varphi.
\end{align}
Following the approach in Ref. \cite{Grenzebach:2014fha}, the tangent vector at the observer can be expressed as
\begin{align}\label{lambda2}
\dot{\lambda}=\gamma(\sin\theta\cos\psi e_1+\sin\theta\sin\psi e_2+\cos\theta e_3-e_0),
\end{align}
where $\gamma$ is a scalar factor and can be solved by substituting Eqs. (\ref{taut}), (\ref{tauvarphi}), (\ref{taur}), and (\ref{tauvartheta}), and Eq. \meq{lambda2} into Eq. \meq{lambda1}. The result gives
\begin{equation}
\gamma=-\frac{k}{\sqrt{\Sigma \Delta_r}},
\end{equation}
and then yield
\begin{align}\label{sintheta}
\sin\theta=&\left.\frac{\sqrt{\eta\Delta_r +\left(a-\chi\xi\right)^2 \Delta_r}}{r^2+a^2-a\chi\xi}\right|_{r=r_0},\\
\label{sinpsi}
\sin\psi=&\left.\frac{\csc\vartheta\left(a\cos^2\vartheta-a+\chi\xi\right)}
{\sqrt{\eta\Delta_\vartheta+\left(a-\chi\xi\right)^2 \Delta_\vartheta}} \right|_{\vartheta=\vartheta_0}.
\end{align}
In order to use analytical parameter to roperesent the boundary curve of the shadow \cite{Grenzebach:2014fha}, we utilized the stereographic projection from the celestial sphere onto a plane, where standard Cartesian coordinates are employed,
\begin{align}\label{dkex}
x(r_p)=&-2\tan\left[\frac{\theta(r_p)}{2}\right]\sin\left[\psi(r_p)\right],\\ \label{dkey}
y(r_p)=&-2\tan\left[\frac{\theta(r_p)}{2}\right]\cos\left[\psi(r_p)\right].
\end{align}
This is a set of parameter equations concerning the radius of the photon sphere $r_p$ \cite{Perlick:2021aok}.

For the non-rotating black hole case with $a=0$, the expression for the impact parameter has a singularity, necessitating separate consideration of the static case. Using the circular orbit condition \meq{YGDTJ} and setting $\eta+\xi^2\chi^2:=1/X$ \cite{Grenzebach:2014fha}, the Eqs. \meq{sintheta}-\meq{sinpsi} have the form:
\begin{align}
\sin\theta^{(a=0)}=&\left.\frac{\sqrt{\Delta_r|_{a=0}}}{r^2\sqrt{X}}\right|_{r=r_0},\\
\sin\psi^{(a=0)}=&\left.\sqrt{X}\chi\xi \csc\vartheta\right|_{\vartheta=\vartheta_0},
\end{align}
where
\begin{align}
X=\frac{1}{2r_p^2}-\frac{(1+\alpha)}{2r_p^3}-\frac{\Lambda}{3},
\end{align}
and the radius of the photon sphere is
\begin{align}
r_p=\frac{3}{2}(1+\alpha)\left(1+\sqrt{1-\frac{8\alpha}{9(1+\alpha)}}\right).
\end{align}
In this case, the system of Eqs. \meq{dkex}-\meq{dkey} will become parameter equations only concerning the impact parameter $\xi$. Note that the expression of the photon sphere radius is independent of $ \Lambda $. From this, one can deduce that the shape of a static black hole is a standard circle, with its radius dependent on the MOG parameter $\alpha$.

\begin{figure}[h]
\centering
\includegraphics[width=0.49\linewidth]{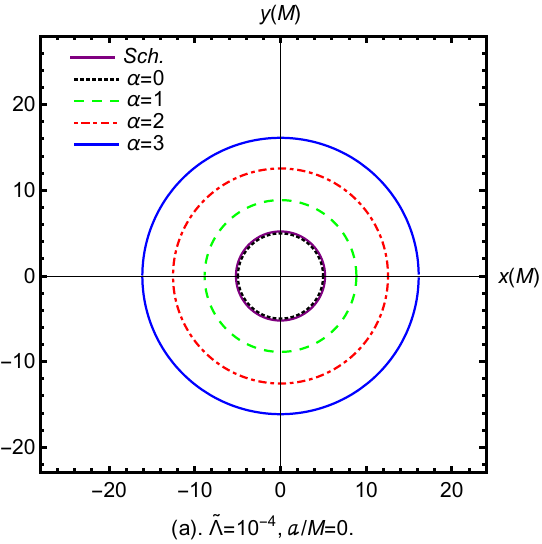}
\includegraphics[width=0.47\linewidth]{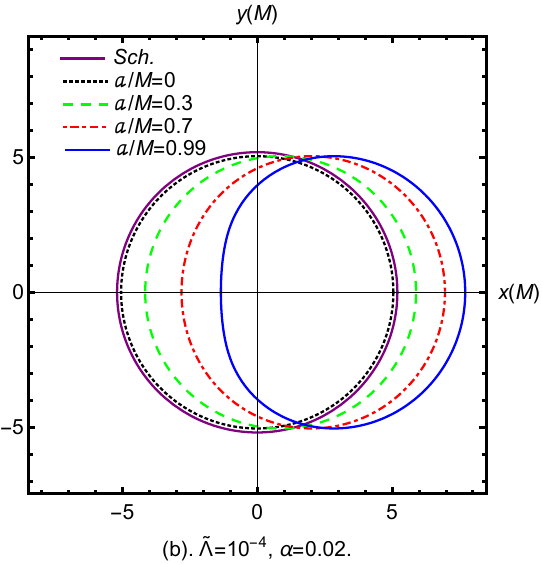}
\includegraphics[width=0.48\linewidth]{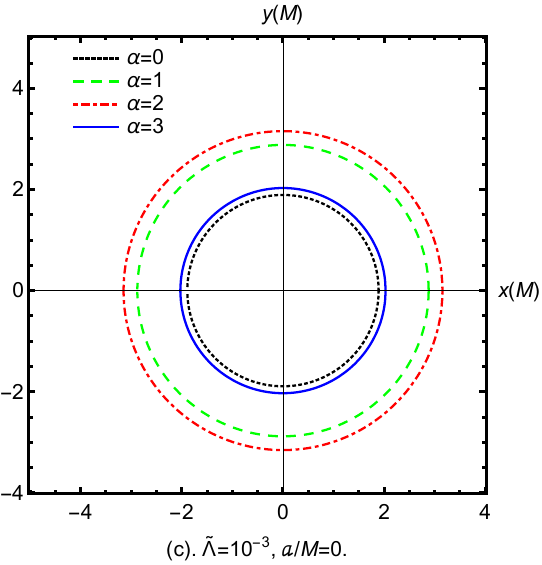}
\includegraphics[width=0.48\linewidth]{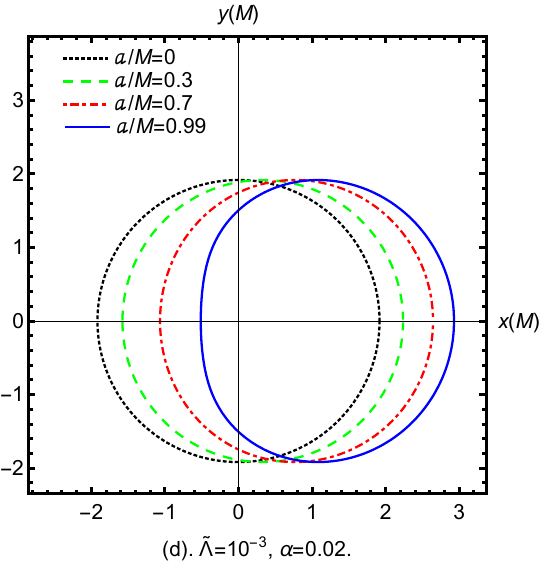}
\caption{Shadows cast by black holes with a cosmological constant, as seen by an observer at $ r_0=50M $ and $ \vartheta_0=\pi/2 $.
The (a) and (c) are static cases $ a/M=0 $, corresponding to $ \tilde{\Lambda}=10^{-4} $ and $ \tilde{\Lambda}=10^{-3} $, respectively.
The (b) and (d) are rotating case $ \alpha=0.02$, corresponding to $ \tilde{\Lambda}=10^{-4} $ and $ \tilde{\Lambda}=10^{-3} $, respectively. }
\label{fig1}
\end{figure}
\begin{figure}[h]
\centering
\includegraphics[width=0.47\linewidth]{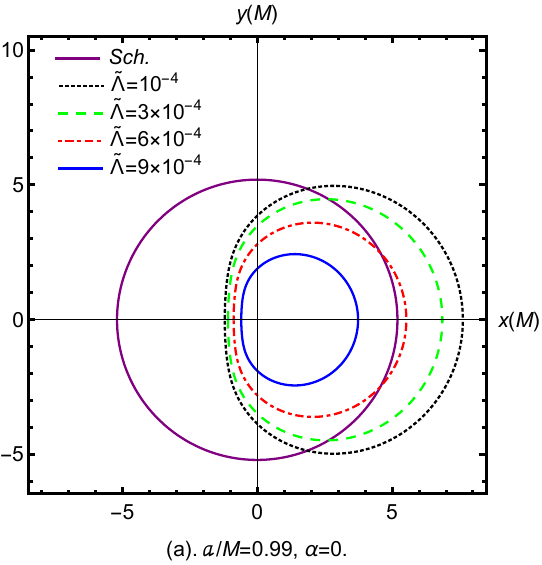}
\includegraphics[width=0.485\linewidth]{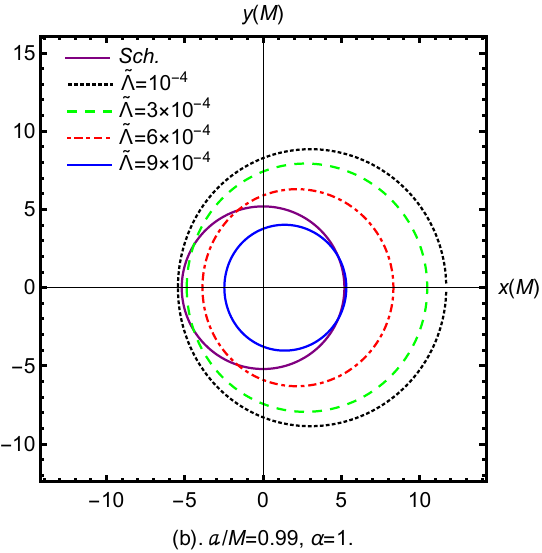}
\includegraphics[width=0.48\linewidth]{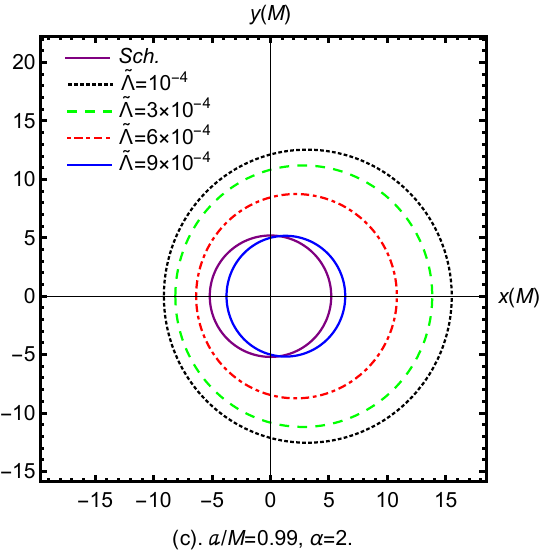}
\includegraphics[width=0.48\linewidth]{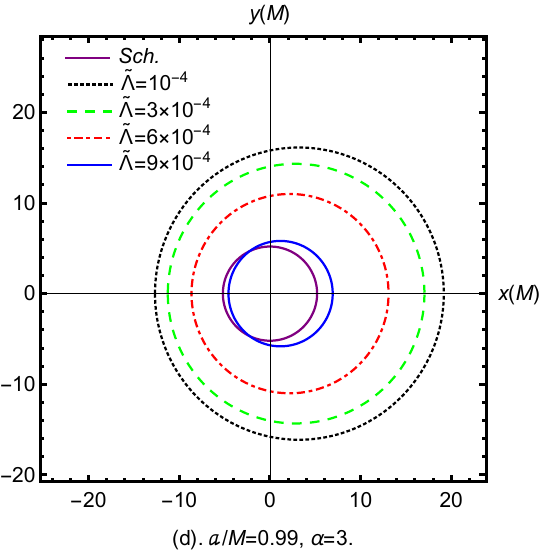}
\caption{
Shadows cast by black holes with a spin $ a/M=0.99 $, as seen by an observer at $ r_0=50M $ and $ \vartheta_0=\pi/2 $.
The (a), (b), (c) and (d) corresponding to $ \alpha=0 $, $ \alpha=1 $, $ \alpha=2 $ and $ \alpha=3 $, respectively.
}
\label{fig2}
\end{figure}
\begin{figure}[h]
\centering
\includegraphics[width=0.48\linewidth]{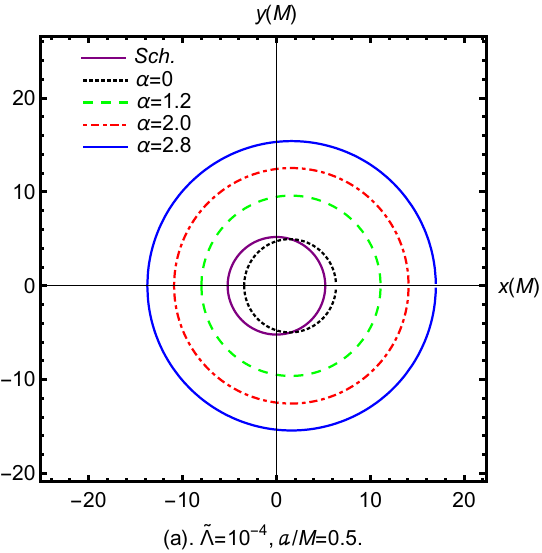}
\includegraphics[width=0.48\linewidth]{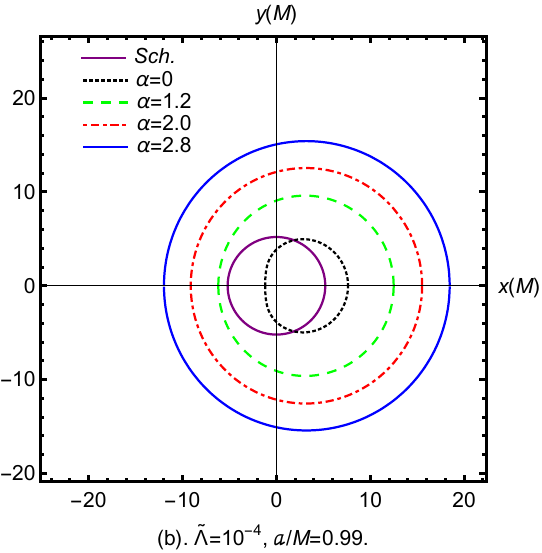}
\includegraphics[width=0.48\linewidth]{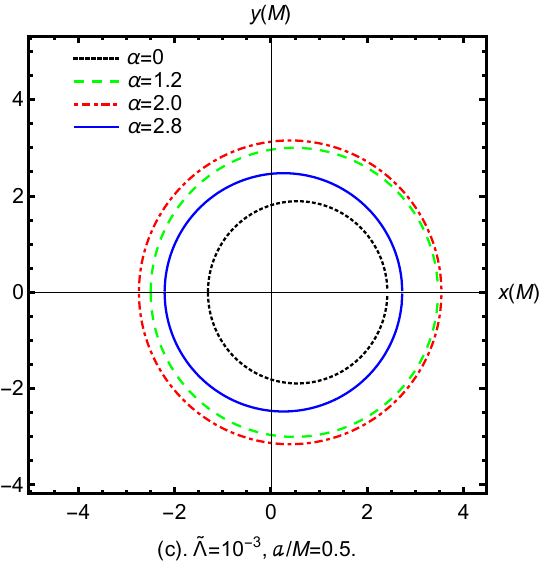}
\includegraphics[width=0.48\linewidth]{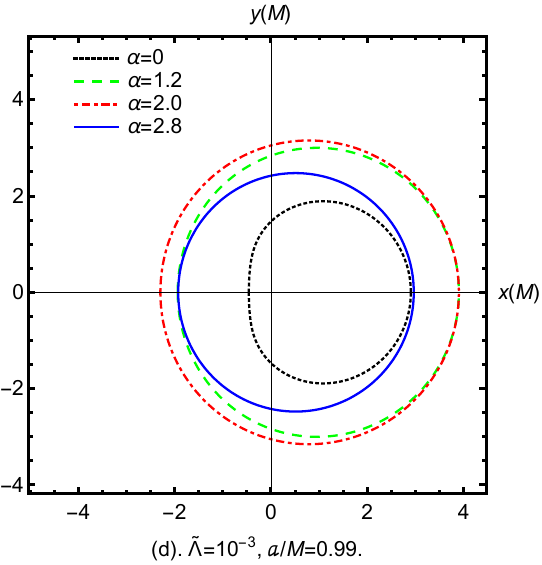}
\caption{
Shadows cast by black holes with a cosmological constant, as seen by an observer at $ r_0=50M $ and $ \vartheta_0=\pi/2 $.
The (a) and (c) are spin $ a/M=0.5 $, corresponding to $ \tilde{\Lambda}=10^{-4} $ and $ \tilde{\Lambda}=10^{-3} $, respectively.
The (b) and (d) are spin $ a/M=0.99$, corresponding to $ \tilde{\Lambda}=10^{-4} $ and $ \tilde{\Lambda}=10^{-3} $, respectively.}
\label{fig3}
\end{figure}
For simplicity, we consider fixing the position of the observer at $ r_0=50M $ and $ \vartheta_0=\pi/2 $. 
Where the angle $ \vartheta_0 $ is defined as the inclination angle between the observer's line of sight and the rotation axis of the black hole. 
From Eqs. \meq{dkex} and \meq{dkey}, one can depict the contour of the black hole shadow under varying MOG parameter $ \alpha $, spin $ a $, or dimensionless cosmological constant $\tilde{\Lambda} $, which is defined as $ \tilde{\Lambda}=\Lambda M^{2} $.
In Fig. \ref{fig1}, the shadows of static black holes and rotating black holes are shown under $ \tilde{\Lambda}=10^{-4} $ or $ \tilde{\Lambda}=10^{-3} $. When the cosmological constant is relatively small, the shadow scale exhibits an increase with the growth of the MOG parameter. 
However, as the cosmological constant increases, this trend changes. 
This shift can be attributed to the observer's position $r_0$ approaching the cosmological horizon $r_c$, causing the observer's position to have a more significant effect on the shadow size than the effects of the MOG modification. 
In Fig. \ref{fig2}, we present the shadow contour of a near-extreme black hole with different $\alpha$ and $\tilde{\Lambda}$, i.e., $a/M=0.99$. 
The results show that, as the MOG parameter increases, the range of the photon sphere radius narrows, and its contour increasingly resembles a standard circle. 
When the cosmological constant is relatively small, the shadow size increases with the growth of the MOG parameter. 
For a fixed $ a $ and $ \alpha $, the apparent shape of the shadow varies with $ \tilde{\Lambda} $. 
As $ \tilde{\Lambda} $ decreases, the shadow's size expands. 
In Fig. \ref{fig3}, the same phenomenon that is clearly visible in Fig. \ref{fig1}-(c) can also be observed in rotating black holes.
Furthermore, when the MOG parameter increases significantly, the rotation-induced distortion of the photon orbits becomes less noticeable. 
It is worth noting that in Figs. \ref{fig1}-\ref{fig3}, the purple solid line represents the shadow of the Schwarzschild black hole, serving as a scale reference.
%\footnote{The coordinates $ x/M $ and $ y/M $ in Figs. \ref{fig1}-\ref{fig3} represent the on-screen scale, which is proportional to the actual celestial coordinate scales $ \bar{x}/M $ and $ \bar{y}/M $ as described by the geometric relationships in equations \meq{dkex} and \meq{dkey}, depending on the observer's position $ r_0 $. For example, in the Schwarzschild case, where $r_0 = 50M$, the relation $\bar{x} \simeq 50.897 x$ holds.\label{ft1}}

The contours of static versus rotating black holes will be extremely similar when the MOG parameter is sufficiently large \cite{Camilloni:2023wyn}, making it difficult to infer black hole parameters solely from the black hole shadow. 
However, does the frame-dragging effect caused by rotation in black hole spacetimes also affect the MOG parameter? 
For this, we can use the numerical ray-tracing method, which adopts the same orthonormal tetrad as the analytical methods in this paper and sets the number of pixels to $ n=2048 $. 
Here, to make the manuscript concise, we would like to show our results immediately; the details of the ray-tracing method are provided in Appendix \ref{AppendixB}.
Figs. \ref{fig4} and \ref{fig5} effectively demonstrate the distortion of space by MOG black holes and the gravitational lensing effects these objects produce. 
Specifically, the gravitational lensing observed can be utilized to examine theories of gravity \cite{Liu:2024bre,AbhishekChowdhuri:2023ekr,Ghosh:2022mka}. 
We can observe that although increasing the MOG parameter makes the contours of rotating black holes approach those of static black holes, it does not affect the frame-dragging effect induced by rotation. 
Moreover, Figs. \ref{fig4} and \ref{fig5} serve as double checks for Figs. \ref{fig1}-(a) and \ref{fig2}-(b), respectively, with each displaying the same black hole contours.
\begin{figure}[h]
\centering
\includegraphics[width=0.45\linewidth]{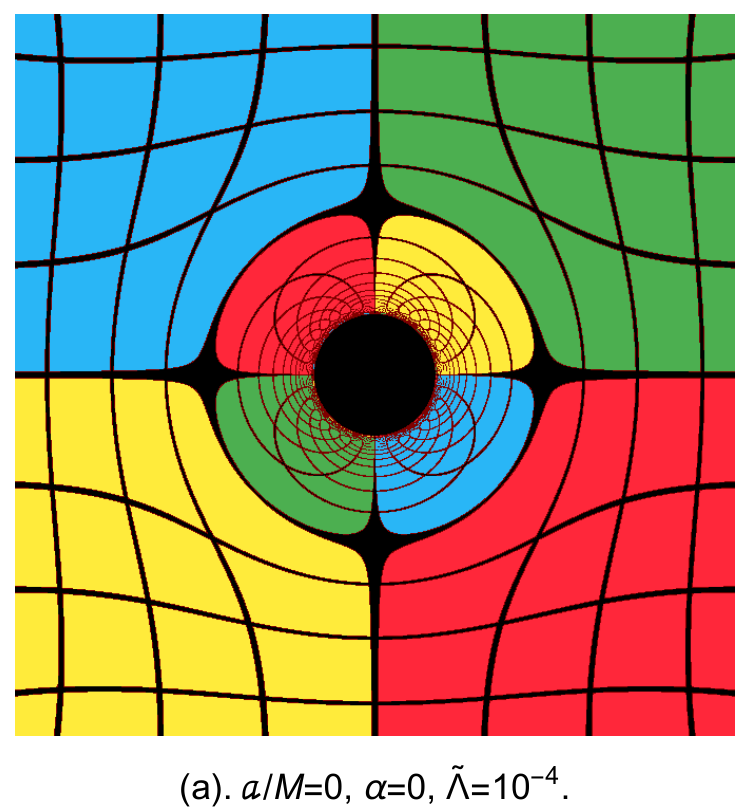}
\includegraphics[width=0.45\linewidth]{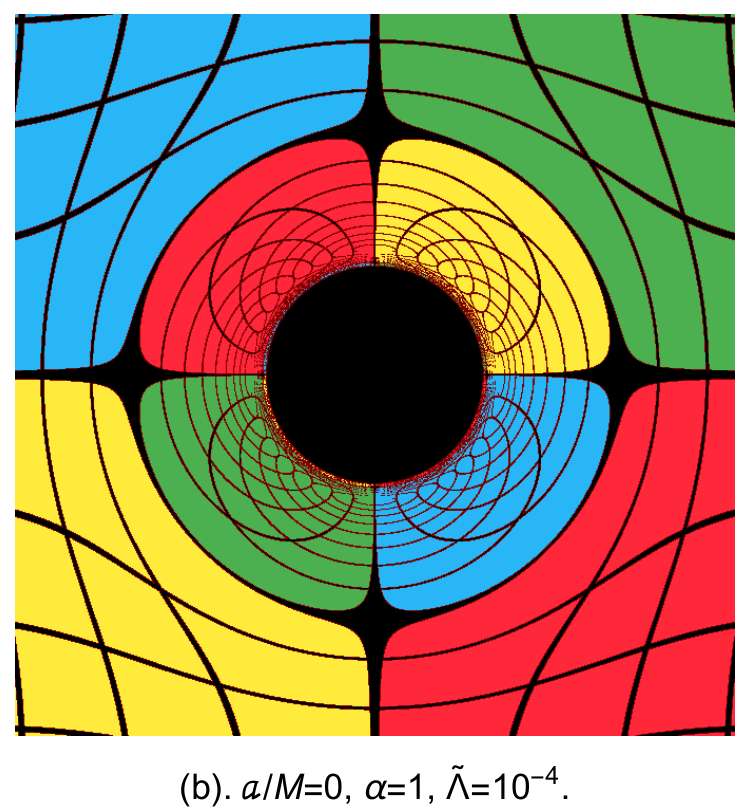}
\includegraphics[width=0.45\linewidth]{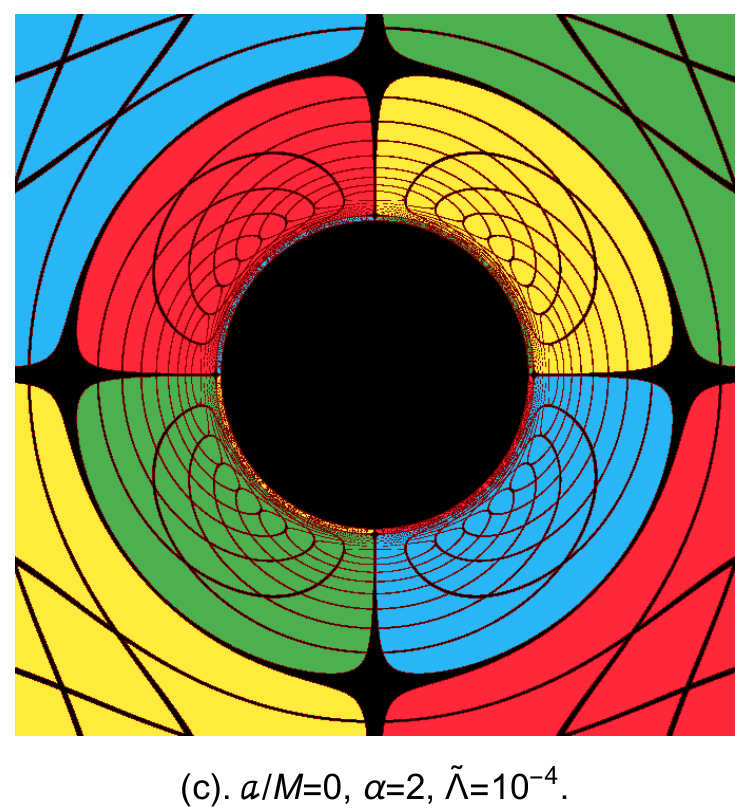}
\includegraphics[width=0.45\linewidth]{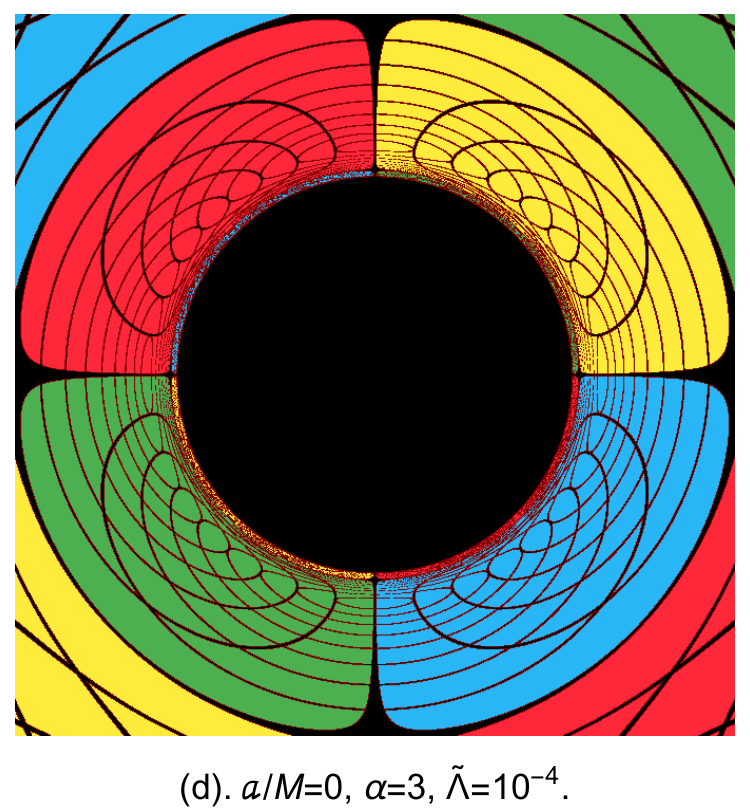}
\caption{Shadows cast by static black holes ($ a/M=0 $) with a cosmological constant, as seen by an observer at $ r_0=50M $ and $ \vartheta_0=\pi/2 $.}
\label{fig4}
\end{figure}
\begin{figure}[h]
\centering
\includegraphics[width=0.45\linewidth]{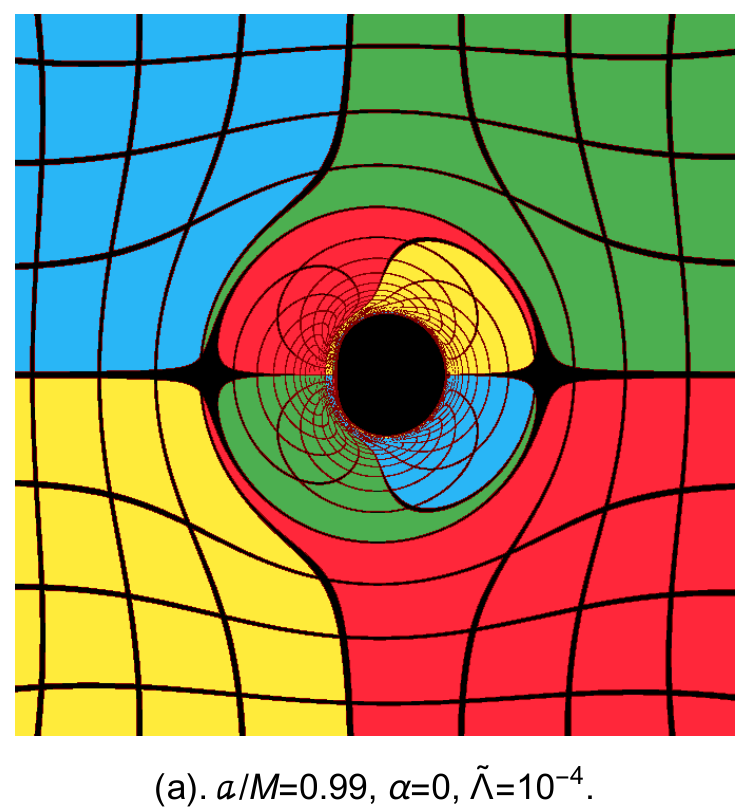}
\includegraphics[width=0.45\linewidth]{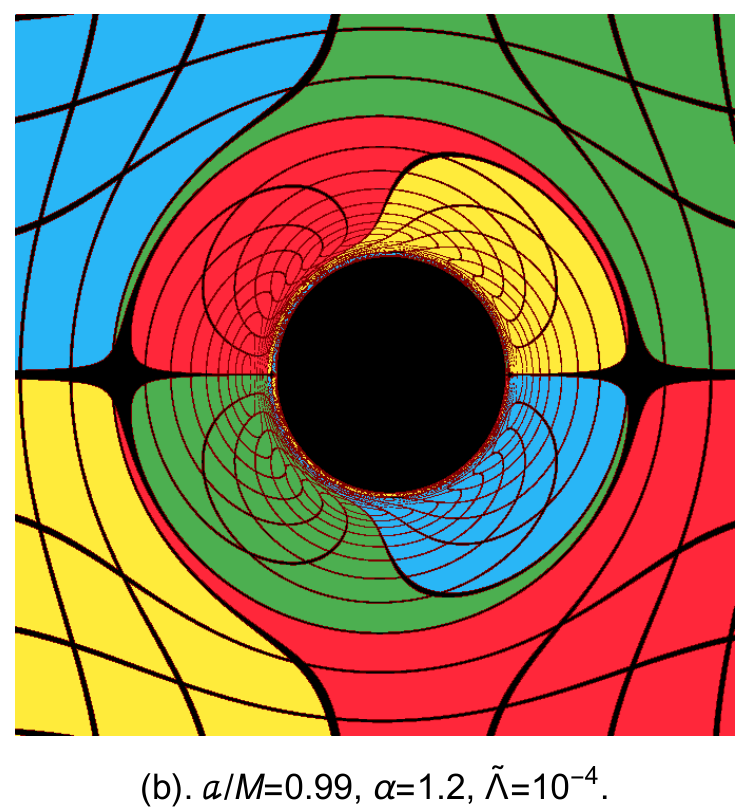}
\includegraphics[width=0.45\linewidth]{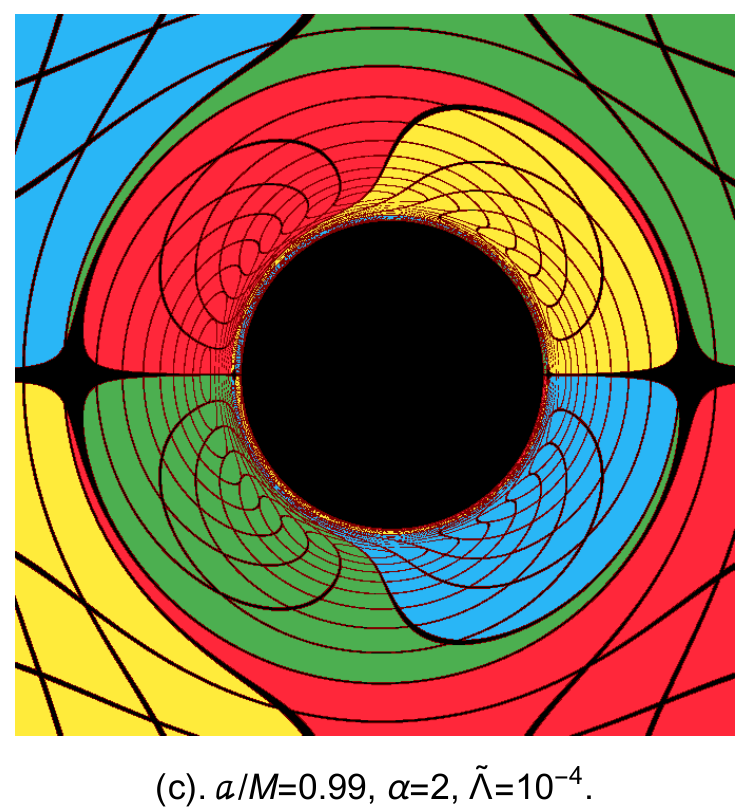}
\includegraphics[width=0.45\linewidth]{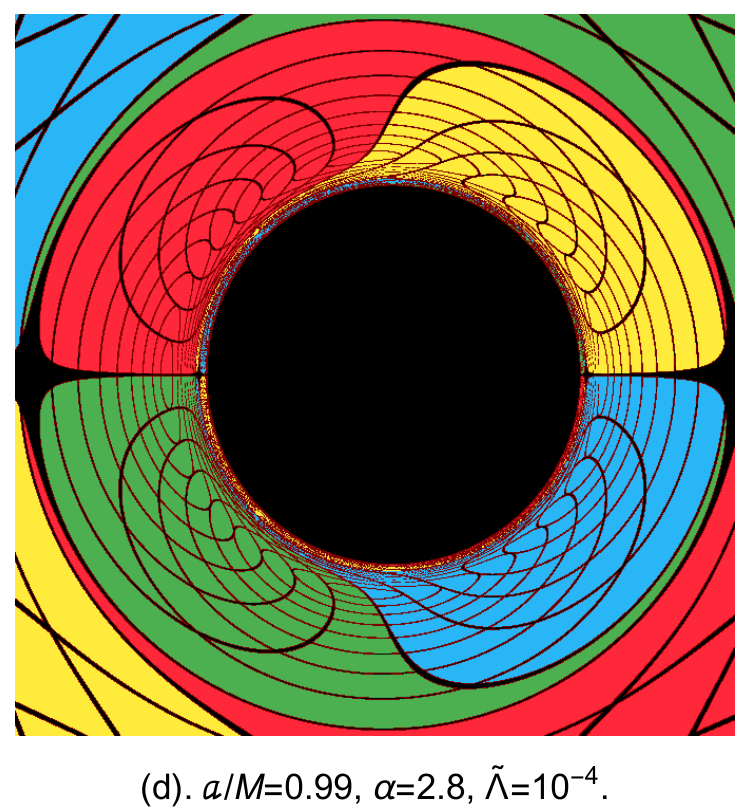}
\caption{Shadows cast by extremely rotating black holes ($ a/M=0.99 $) with a cosmological constant, as seen by an observer at $ r_0=50M $ and $ \vartheta_0=\pi/2 $.}
\label{fig5}
\end{figure}

%%%%%%%%%%%%%%%%%%%%%%%%%%%%%%%%%%%%%%%%%%%%%%%%%%%%
\subsection{Observable measurements}\label{Sec.4.2}
%%%%%%%%%%%%%%%%%%%%%%%%%%%%%%%%%%%%%%%%%%%%%%%%%%%%

Up to this point, we have determined that parameters such as spin $a$, MOG parameter $\alpha$, and dimensionless cosmological constant $\tilde{\Lambda}$ will significantly affect the apparent shape of the shadow in Kerr-MOG-dS spacetime. 
To characterize the shadow, it is essential to have easily measurable and reliable properties of observable measurements. 
We adopt two such observable measurements: the radius $ R_s $ and distortion parameter $ \delta_s $, following an approach analogous to the one taken in Ref. \cite{Hioki:2009na}.
\begin{figure}[ht]
\centering
\includegraphics[width=0.7\linewidth]{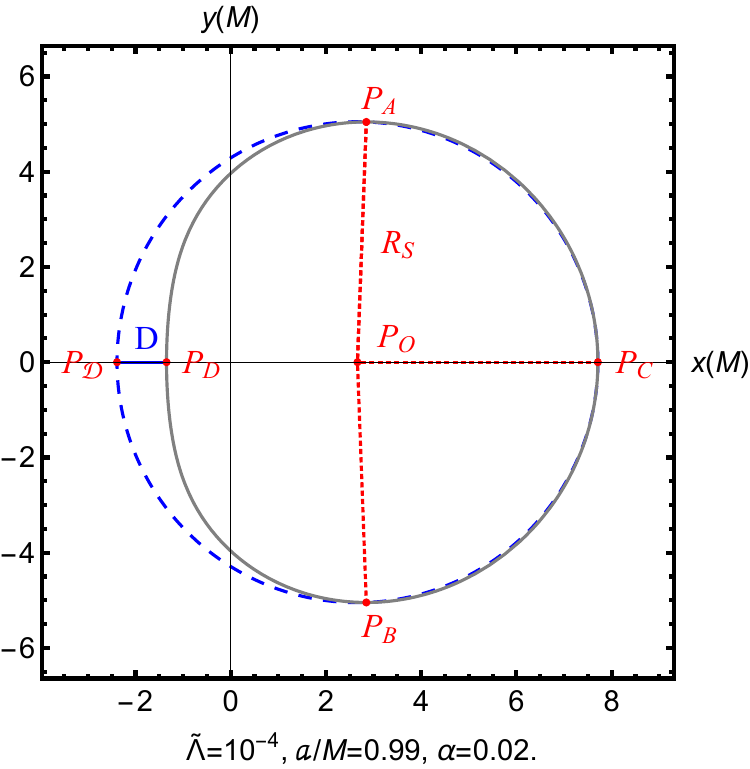}
\caption{Shadows cast by black holes, as seen by an observer at $ r_0=50M $ and $ \vartheta_0=\pi/2 $. The observable measurements for the apparent shape of the black hole are the radius of the reference circle $ R_s $ and the distortion parameter $ \delta_s=\mathrm{D}/R_s $, where $ \mathrm{D} $ is the difference between the left endpoints of the reference circle and of the shadow.}
\label{fig6}
\end{figure}
The parameter $ R_s $ characterizes the size of the shadow, corresponding to the radius of the reference circle depicted as a blue dashed line in Fig. \ref{fig6}. The circle passes through three points: the top position $P_A=(x_t,y_t)$, the bottom position $P_B=(x_b,y_b)$, and the point $P_C=(x_r, 0)$, which corresponds to the unstable retrograde circular orbit as observed from the equatorial plane by an observer. Additionally, the point $ P_O=(x_o,0) $ represents the center of the reference circle, and $ x_o $ can be obtained from the coordinates of $ P_A, P_B $, and $ P_C $:
\begin{align}
x_o=\frac{x^2_r-x^2_t-y^2_t}{2(x_r-x_t)}.
\end{align}
Afterwards, the difference between the shaded left-hand point $ P_D=(x_d,0) $ and the reference circle left-hand point $ P_\mathcal{D}=(\tilde{x}_r,0) $ needs to be taken into account, with the size of this difference being evaluated by $ \mathrm{D}=|x_d-\tilde{x}_r|$. Further, the two observable measurements take the form
\begin{align}
R_s=&\frac{(x_t-x_r)^2+y_t^2}{2|x_r-x_t|}, \\ \delta_s=&\frac{\mathrm{D}}{R_s}.
\end{align}

\begin{figure}[h]
\centering
\includegraphics[width=0.47\linewidth]{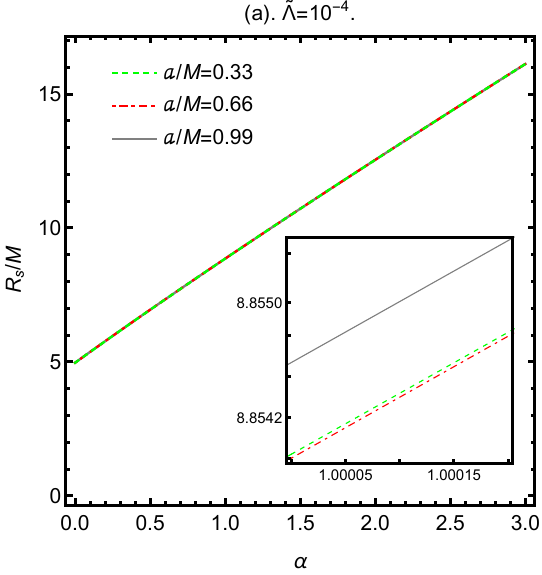}
\includegraphics[width=0.485\linewidth]{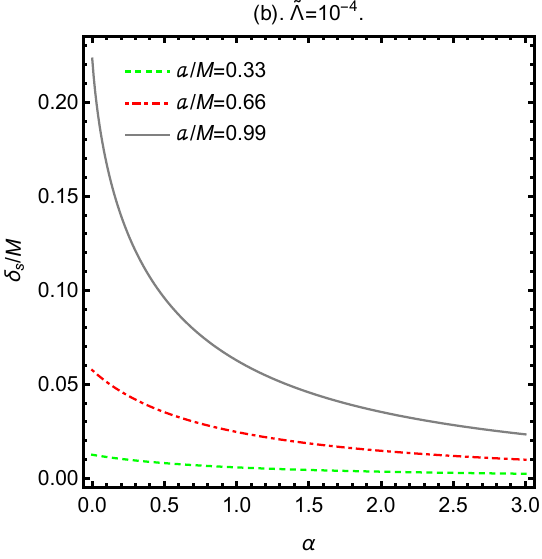}
\includegraphics[width=0.475\linewidth]{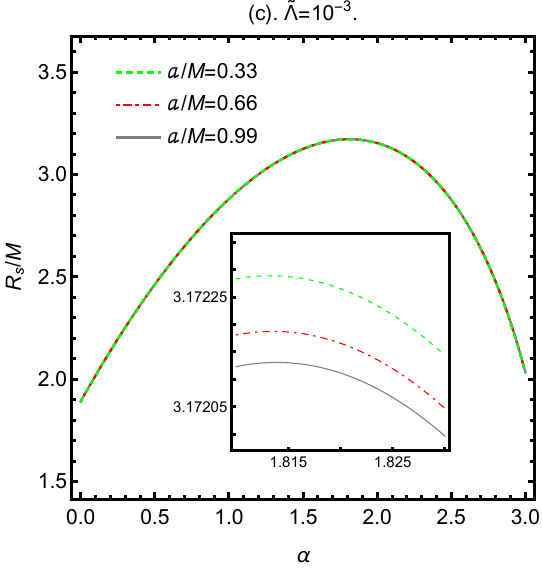}
\includegraphics[width=0.48\linewidth]{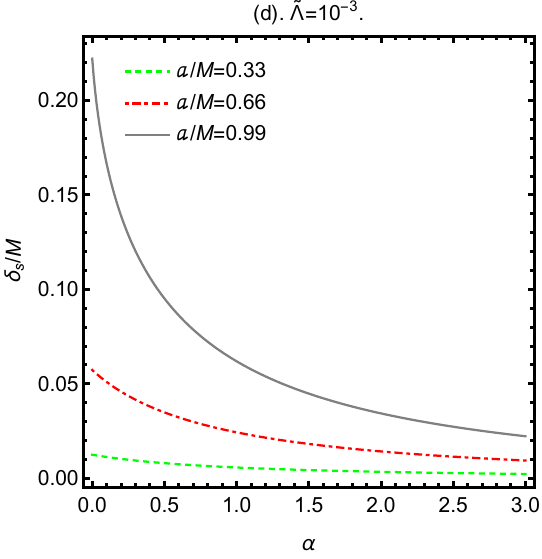}
\caption{Observable measurements $R_s$ (left) and $\delta_s$ (right) as functions of $\alpha$ are presented, respectively, for $\tilde{\Lambda}=10^{-4}$ (top) and $\tilde{\Lambda}=10^{-3}$ (bottom). The observer is placed at $ r_0 = 50M $ and $ \vartheta_0=\pi/2. $ }
\label{fig7}
\end{figure}
We numerically calculated these two observable measurements under the condition of fixing the observer position at $r_0 = 50M$ with an inclination angle of $\vartheta_0= \pi/2$. The results are presented in Fig. \ref{fig7}. Note that the static case is not included in the results because the contour of the shadow coincides with the reference circle, resulting in no distortion. Figs. \ref{fig7}-(b) and \ref{fig7}-(d) show that the cosmological constant $\tilde{\Lambda}$ has little effect on $\delta_s$. But $\delta_s$ decreases as the parameter $\alpha$ increases, which suggests that the Kerr-MOG-dS black hole gets less deformed compared to the Kerr-dS black hole. Figs. \ref{fig7}-(a) and \ref{fig7}-(c) clearly show the radius $R_s$ of the shadow. From these results, it becomes evident that when the cosmological constant $ \tilde{\Lambda}=10^{-4} $, $ R_s $ increases as the parameter $ \alpha $ increases with $ \alpha \in [0,3] $. 
However, when the cosmological constant $ \tilde{\Lambda}=10^{-3} $, there is a maximum value for $ R_s $ , corresponding to $ \alpha \approx 1.815 $.
This is consistent with the peculiar phenomena observed in Figs. \ref{fig1}-(c), \ref{fig3}-(c), and \ref{fig3}-(d) describing the shadow cast by MOG-dS black holes in Sec. \ref{Sec.4}.
\begin{figure}[h]
\centering
\includegraphics[width=0.48\linewidth]{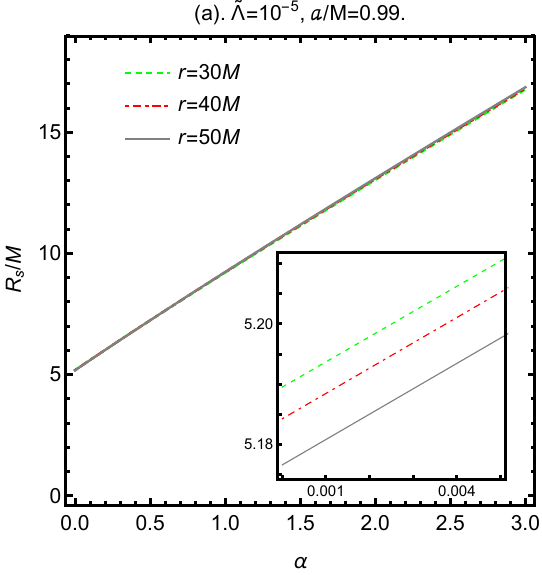}
\includegraphics[width=0.48\linewidth]{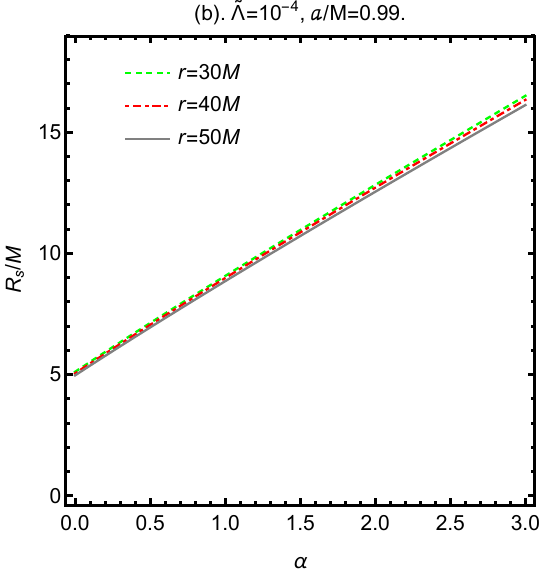}
\includegraphics[width=0.48\linewidth]{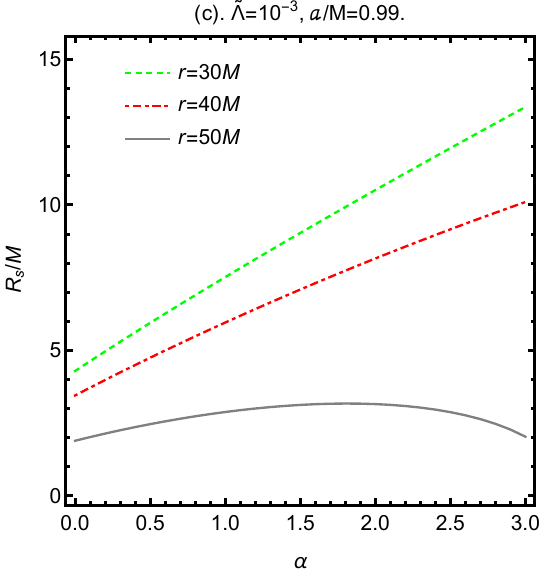}
\includegraphics[width=0.48\linewidth]{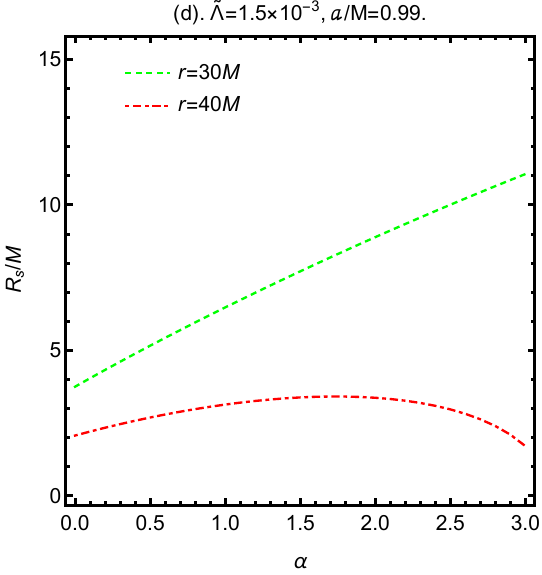}
\caption{
Observable measurements $R_s$ as functions of $\alpha$ are presented, with the positions of the observers located at $r_0=30M$, $r_0=40M$, and $r_0=50M$, and an observation angle of $\vartheta_0=\pi/2$.}
\label{fig8}
\end{figure}
To understand this phenomenon, we explored the impact of the observer's position on $ R_s $. 
For smaller cosmological constants, the domain of outer communication is immensely vast, such that the observer's position has a negligible effect on $ R_s $, as shown in Figs. \ref{fig8}-(a) and \ref{fig8}-(b).
However, for larger cosmological constant, the domain of outer communication is more limited and further decreases with increasing MOG parameters. 
When the observer's position is close to the de Sitter horizon, $R_s$ rapidly decreases, as shown in Fig. \ref{fig8}-(c) with $r=50M$ and Fig. \ref{fig8}-(d) with $r=40M$. 
The corresponding cosmological horizons are located at points from $53.744M$ to $50.378M$ with $\alpha \in [0,3]$, and from $43.686M$ to $40.214M$ with $\alpha \in [0,3]$, respectively.

For a approximatively estimation, utilizing the Kerr-MOG-dS metric \meq{ds2MOG2}, we calculate the angular radius of a black hole shadow, defined as $ \theta_{\text{BH}} = R_s \frac{\mathcal{M}}{D_O} $, with $D_O$ representing the distance from the observer to the black hole. 
%The radius $ \bar{R}_s $, which is proportional to the on-screen radius $ R_s $ as described by the geometric relationships in equations \meq{dkex} and \meq{dkey} (see detail in footnote \ref{ft1}). 
Specifically, for a black hole possessing a mass of $ \mathcal{M} $ and situated $ D_O $ away from the observer, we express the angular radius $ \theta_\text{BH} $ quantitatively as $ \theta_\text{BH} = 9.87098 \times 10^{-6} R_s \left(\frac{\mathcal{M}}{M_\odot}\right) \left(\frac{1 \text{kpc}}{D_O}\right) \mu\text{as} $ \cite{Amarilla:2011fx}.

In Tables \ref{tab1} and \ref{tab2}, we present the calculated angular radius of the black holes at the Galactic center, Sgr A*, and M87*, respectively, using the Kerr-MOG-dS metric. Here, we utilize the latest observations which indicate that the mass of the black hole Sgr A* is $\mathcal{M} = 4.0 \times 10^6 M_\odot$ with an observer distance of $D_O = 8.3 kpc$ \cite{EventHorizonTelescope:2022wkp}, and for the black hole M87*, the mass is $\mathcal{M} = 6.5 \times 10^9 M_\odot$ with an observer distance of $D_O = 16.8 Mpc$ \cite{EventHorizonTelescope:2019ggy}. It should be noted that we have highlighted the data corresponding to the observed range of angular diameters for the Sgr A* and M87* black holes from recent observations.\footnote{The angular diameters of the Sgr A* and M87* black holes are measured as $51.8 \pm 2.3 \mu\text{as}$ \cite{EventHorizonTelescope:2022wkp} and $42 \pm 3.0 \mu\text{as}$ \cite{EventHorizonTelescope:2019dse,Long:2019nox,Pal:2023wqg}, respectively.} By combining it with the data in Tables \ref{tab1} and \ref{tab2}, one can find that there is room for the theoretical model of such a Kerr-MOG-dS black hole. Furthermore, the inclusion of the cosmological constant provides a larger parameter space for the MOG parameters.
\begin{table}[h]
\renewcommand{\arraystretch}{1.25}
\centering
\setlength\tabcolsep{1.3mm}{
\begin{tabular}{ccccccccc}
\hline\hline
\multirow{2}{*}{$ \theta_\text{BH}(\mu arcsec) $} & \multirow{2}{*}{$\alpha=0$}& \multirow{2}{*}{$\alpha=0.1$}& \multirow{2}{*}{$\alpha=0.2$}& \multirow{2}{*}{$\alpha=0.3$}& \multirow{2}{*}{$\alpha=0.4$}
\\ \\
\hline
  $\tilde{\Lambda}=0.0000 $  &24.737          &\textbf{26.747}   &28.738           &30.712             &32.670   \\
  $\tilde{\Lambda}=0.0001 $  &23.644          &\textbf{25.562}   &27.460           &29.342             &31.208   \\
  $\tilde{\Lambda}=0.0002 $  &22.497          &24.317            &\textbf{26.118}  &27.902             &29.671   \\
  $\tilde{\Lambda}=0.0003 $  &21.288          &23.004            &24.701           &\textbf{26.382}    &28.047   \\
\\
  $\tilde{\Lambda}=0.0000 $  &\textbf{24.751} &\textbf{26.760}  &28.750            &30.722             &32.679    \\
  $\tilde{\Lambda}=0.0001 $  &23.658          &\textbf{25.573}  &27.470            &29.351             &31.215    \\
  $\tilde{\Lambda}=0.0002 $  &22.510          &24.328           &\textbf{26.127}   &27.909             &29.677    \\
  $\tilde{\Lambda}=0.0003 $  &21.298          &23.013           &24.709            &\textbf{26.388}    &28.053    \\
\hline\hline
\end{tabular}}
\caption{The numerical estimation of the angular radius of the supermassive black hole Sgr A* in our galaxy using the metric of a Kerr-MOG-dS black hole.  The top four rows correspond to $ a/M=0.1 $; the bottom four rows to $ a/M=0.99 $.}
\label{tab1}
\end{table}
\begin{table}[h]
\renewcommand{\arraystretch}{1.25}
\centering
\setlength\tabcolsep{1.3mm}{
\begin{tabular}{ccccccccc}
\hline\hline
\multirow{2}{*}{$ \theta_\text{BH}(\mu arcsec) $} & \multirow{2}{*}{$\alpha=0$}& \multirow{2}{*}{$\alpha=0.1$}& \multirow{2}{*}{$\alpha=0.2$}& \multirow{2}{*}{$\alpha=0.3$}& \multirow{2}{*}{$\alpha=0.4$}
\\ \\
\hline
  $\tilde{\Lambda}=0.0000 $  &\textbf{19.859}   &\textbf{21.473}  &23.081             &24.656            &26.229     \\
  $\tilde{\Lambda}=0.0001 $  &18.982            &\textbf{20.522}  &\textbf{22.046}    &23.556            &25.055     \\
  $\tilde{\Lambda}=0.0002 $  &18.062            &\textbf{19.522}  &\textbf{20.968}    &\textbf{22.401}   &23.821     \\
  $\tilde{\Lambda}=0.0003 $  &17.090            &18.457           &\textbf{19.816}    &\textbf{21.164}   &22.517     \\
\\
  $\tilde{\Lambda}=0.0000 $  &\textbf{19.871}  &\textbf{21.484}   &23.064             &24.664            &26.236     \\
  $\tilde{\Lambda}=0.0001 $  &18.993           &\textbf{20.531}   &\textbf{22.054}    &23.563            &25.061     \\
  $\tilde{\Lambda}=0.0002 $  &18.071           &\textbf{19.531}   &\textbf{20.975}    &\textbf{22.406}   &23.826     \\
  $\tilde{\Lambda}=0.0003 $  &17.099           &18.475            &\textbf{19.837}    &\textbf{21.185}   &22.521     \\
\hline\hline
\end{tabular}}
\caption{The numerical estimation of the angular radius of the supermassive black hole M87* using the metric of a Kerr-MOG-dS black hole.  The top four rows correspond to $ a/M=0.1 $; the bottom four rows to $ a/M=0.99 $.}
\label{tab2}
\end{table}

%%%%%%%%%%%%%%%%%%%%%%%%%%%%%%%%%%%%%%%%%%%%%%%
\section{CONCLUSIONS AND OUTLOOKS}\label{Sec.5}
%%%%%%%%%%%%%%%%%%%%%%%%%%%%%%%%%%%%%%%%%%%%%%%

In this work, we have extend the Kerr-MOG black hole solution \cite{Moffat:2014aja} to more general cases with a cosmological constant (namely, the Kerr-MOG-dS and Kerr-MOG-AdS black hole solutions) and investigate the black hole shadows of the Kerr-MOG-dS case. Due to the presence of the positive constant $\Lambda$, the observer should be positioned between the black hole's event horizon and the cosmological horizon, rather than at infinity. The apparent shape of the shadow depends on the observer's position.

A summary of three novel consequences of the shadows of the Kerr-MOG-dS black hole obtained in this paper are listed in the following order:
\begin{itemize}[leftmargin=*,noitemsep,topsep=0pt,partopsep=0pt]
\item \textbf{Size:}
As the cosmological constant increases, the size of the black hole shadow decreases. But the size of the shadow does not appear to be monotonic as $\alpha$ changes. For example, it would reach a maximum at $\alpha\approx 1.815$ with $\tilde{\Lambda}=10^{-3}$.

\item \textbf{Shape:}
The shape of the shadow becomes more rounded as the MOG parameter $\alpha$ increases, causing the contours of rotating black holes to approach those of static black holes. However, this change does not impact the frame-dragging effect induced by rotation. The distortion of space is consistent with that observed in GR.

\item \textbf{Observable measurements:}
For observable measurements, we consider the radius $R_s$, which is associated with the apparent size, and the distortions $\delta_s$, which relate to the deformation of the shadow. The results show that, with the position of the observer held constant and the cosmological constant sufficiently large, the radius $R_s$ reaches a maximum value as $\alpha$ increases. This increase in $\alpha$ concurrently reduces the distortion $\delta_s$.
\end{itemize}
Our results indicate that the interplay between the cosmological constant and MOG parameter leads to intriguing phenomena in the black hole shadow contours.
If the Earth were situated at the edge of the universe in this black hole spacetime, the black hole spin parameter and MOG parameter would become degenerate.
Additionally, we used the latest observational data from M87* and Sgr A* to impose certain parameter constraints on Kerr-MOG-dS black holes, confirming the validity of STVG theory.

There are two promising further topics to be pursued in the future. An intriguing topic is to explore other observable measurements of the shadow of the Kerr-MOG-dS black hole. To prevent certain shadow degeneracy caused by black hole parameters, we can explore other distortion parameters, such as the oblateness parameter $K_s$, and the "thickness" parameter $T$ of the shadow \cite{Wang:2018eui}. In order to visualize the shadow in a real astronomical setting, one should rely on the emissions originating from the accretion disk surrounding the black hole \cite{Zhang:2021hit,Zhang:2022klr}. And the other is to investigate the properties of the Kerr-MOG-AdS black hole, such as black hole thermodynamics \cite{Christodoulou:1970wf,Bardeen:1973gs,Wu:2019pzr,Cong:2021fnf,Ahmed:2023snm,Liu:2024oas}, thermodynamic topological classification \cite{Wei:2022dzw,Wu:2022whe,Wu:2023sue,Wu:2023xpq}, null hypersurface caustics \cite{AlBalushi:2019obu,Imseis:2020vsw,
Wu:2021pfp}, phase transition criticality \cite{Kubiznak:2012wp,Ahmed:2023dnh,Wu:2024rmv}, Joule-Thomson expansion \cite{Okcu:2017qgo}, etc.

\acknowledgments
The authors sincerely thank Professor Songbai Chen for his contributions to this subject and Dr. Fen Long for the profound discussions. 
This work was supported by the National Natural Science Foundation of China under Grants No. 12122504, No. 12375046, No. 12035005, No. 12205243, No. 12375053, by the innovative research group of Hunan Province under Grant No. 2024JJ1006, by the Natural Science Foundation of Hunan Province under grant No. 2023JJ30384, by the Hunan provincial major sci-tech program under grant No.2023zk1010, by the Sichuan Science and Technology Program under Grant No. 2023NSFSC1347, and by the Doctoral Research Initiation Project of China West Normal University under Grant No. 21E028.

\appendix
\section{Energy-Momentum Tensor}\label{AppendixA}
In this appendix, we provide the energy-momentum tensor corresponding to the metric \meq{dsKS3}. The MOG vector potential \meq{MOGfield} in the coordinates $(\bar{t}, \bar{r}, \bar{\vartheta}, \bar{\varphi})$ is given as follows:
\begin{equation}
\begin{aligned}
\bar{\phi}=&-\frac{\sqrt{G_\text{N}\alpha}M\bar{r}}{\Sigma}\left(\frac{\sin^2\bar{\vartheta}}{\chi}+\cos^2\bar{\vartheta}\right)d\bar{t}\\
&-\frac{h\sqrt{G_\text{N}\alpha}M\bar{r}}{\Delta_r\Delta_{\bar{r}}}d\bar{r}+\frac{\sqrt{G_\text{N}\alpha}M\bar{r}}{\chi\Sigma}a\sin^2\bar{\vartheta}d\bar{\varphi},
\end{aligned}
\end{equation}
All non-zero components of $ T_{ab} $ are then given as follows:
\begin{equation}
\begin{aligned}
T_{00}=&-\frac{1}{24\pi \chi^2 \Sigma^3}G_\text{N}M^2\alpha\Delta_{\bar{\vartheta}}
\left[(3-\bar{r}^2\Lambda)\Sigma_2+3h\Delta_{\bar{\vartheta}}\right],\\
T_{01}=&-\frac{1}{8\pi\chi\Delta_{\bar{r}}\Sigma^2}G_\text{N}M^2\alpha h\Delta_{\bar{\vartheta}},\\
T_{11}=&-\frac{1}{8\pi\Delta_{\bar{r}}^2\Sigma}G_\text{N}M^2\alpha\left(\Delta_r-2\Delta_{\bar{r}}\right),\\
T_{03}=&\frac{1}{8\pi \chi^2 \Sigma^3} G_\text{N}M^2\alpha\Delta_{\bar{\vartheta}}
\left(\Delta_r+\Delta_{\bar{r}}\right)a\sin^2\bar{{\vartheta}},\\
T_{22}=&-\frac{1}{8\pi\Delta_{\bar{\vartheta}}\Sigma}G_\text{N}M^2\alpha,\\ T_{13}=&\frac{1}{8\pi\chi\Delta_{\bar{r}}\Sigma^2}G_\text{N}M^2\alpha ha\sin^2\bar{\vartheta},\\
T_{33}=&-\frac{\sin^2\bar{{\vartheta}}}{8\pi\chi^2\Sigma^3}G_\text{N}M^2\alpha\left[(\bar{r}^2+a^2)\Sigma_2+ha^2\sin^2\bar{\vartheta}\right].
\end{aligned}
\end{equation}
This set of equations corresponds one-to-one with the effective tensor component equations \meq{E00}. It is clear that this will lead to the effective tensor $E_{ab} = 0$, ensuring that the solutions satisfy the field equations.

\section{Ray-Tracing Method}\label{AppendixB}
In this appendix, we aim to provide a concise and useful overview of the numerical ray-tracing method used in our manuscript.

For Kerr-MOG-dS spacetime, the 4-momentum vector of a photon is written as
\begin{align}
p^a=(\dot{t},\dot{r},\dot{\vartheta},\dot{\varphi}),
\end{align}
where the dot denotes the derivative with respect to the affine parameter $ \tau $.
Given that $ g_{ab}p^{a}p^{b}=0 $, we can define a dual vector
\begin{align}
q_a=g_{ab}p^b,
\end{align}
which satisfies $ q_a p^a=0 $.
The geodesic equations, in the Hamiltonian canonical formalism, are given by
\begin{align}
\dot{q}_a=-\frac{\pp H}{\pp x^a},&&\dot{x}^a=\frac{\pp H}{\pp q_a}
\end{align} 
where $ H $ is the Hamiltonian of the null particle and $ x^a $ are the 4-positions of the geodesic.
These are first-order differential equations with respect to the canonical variables $ (x^a,q_a) $, making them very convenient for numerical geodesic evolution.

\begin{figure}[h]
\centering
\includegraphics[width=0.9\linewidth]{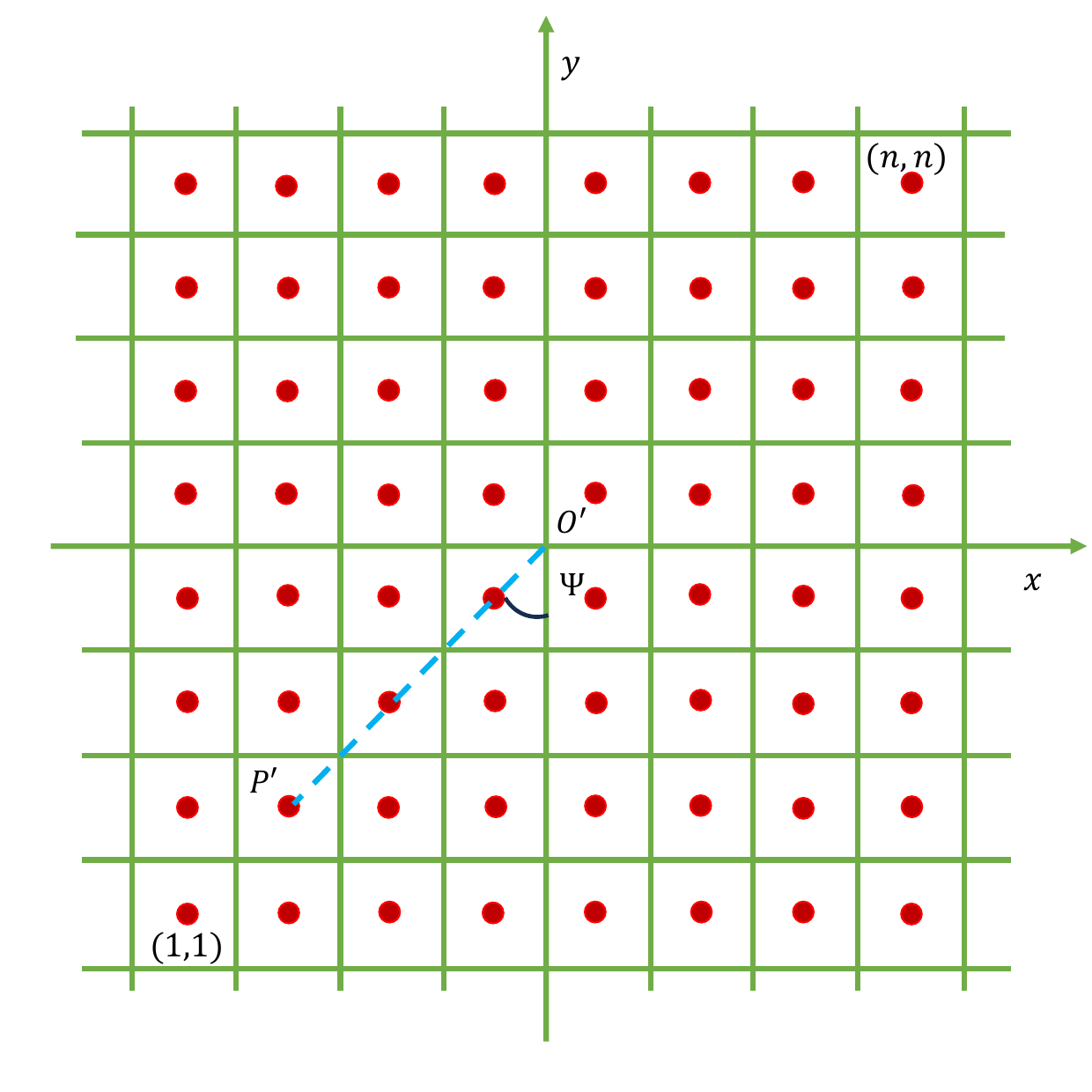}
\caption{
Illustration of the pixels. 
A standard Cartesian coordinate with the origin $ O' $ is set up.}
\label{fig9}
\end{figure}

Just like the approach in Sec. \ref{4.1}, we assume that an observer is located at $ {r_0,\vartheta_0} $ in the coordinates $ \{ t,r,\vartheta,\varphi \} $ and has chosen an orthonormal tetrad the same as in equation \meq{zjbj}.
In the observer's frame, the tangent vector of the null geodesic can be expressed as we can rewrite equation \meq{lambda2} in this form,
\begin{equation*}
\dot{\lambda}=|\overrightarrow{OP}|\gamma(\sin\theta\cos\psi e_1+\sin\theta\sin\psi e_2+\cos\theta e_3- e_0),
\end{equation*}
where $ |\overrightarrow{OP}| $ denotes the tangent vector of the null geodesic at the point $ O $ in the three-dimensional space.
Then, we can re-obtain the stereographic projection from the celestial sphere onto a plane
\begin{equation}\label{B4}
\begin{aligned}
x_{P'}=&-2|\overrightarrow{OP}|\tan\frac{\theta}{2}\sin\psi,\\ 
y_{P'}=&-2|\overrightarrow{OP}|\tan\frac{\theta}{2}\cos\psi,
\end{aligned}
\end{equation}
$ P' $ represents the projection of point $ P $ in the celestial coordinate system onto the Cartesian coordinate system.
We can set up standard Cartesian coordinates with the origin $ O' $ , as shown in Fig. \ref{fig9}.

\begin{figure}[h]
\centering
\includegraphics[width=0.9\linewidth]{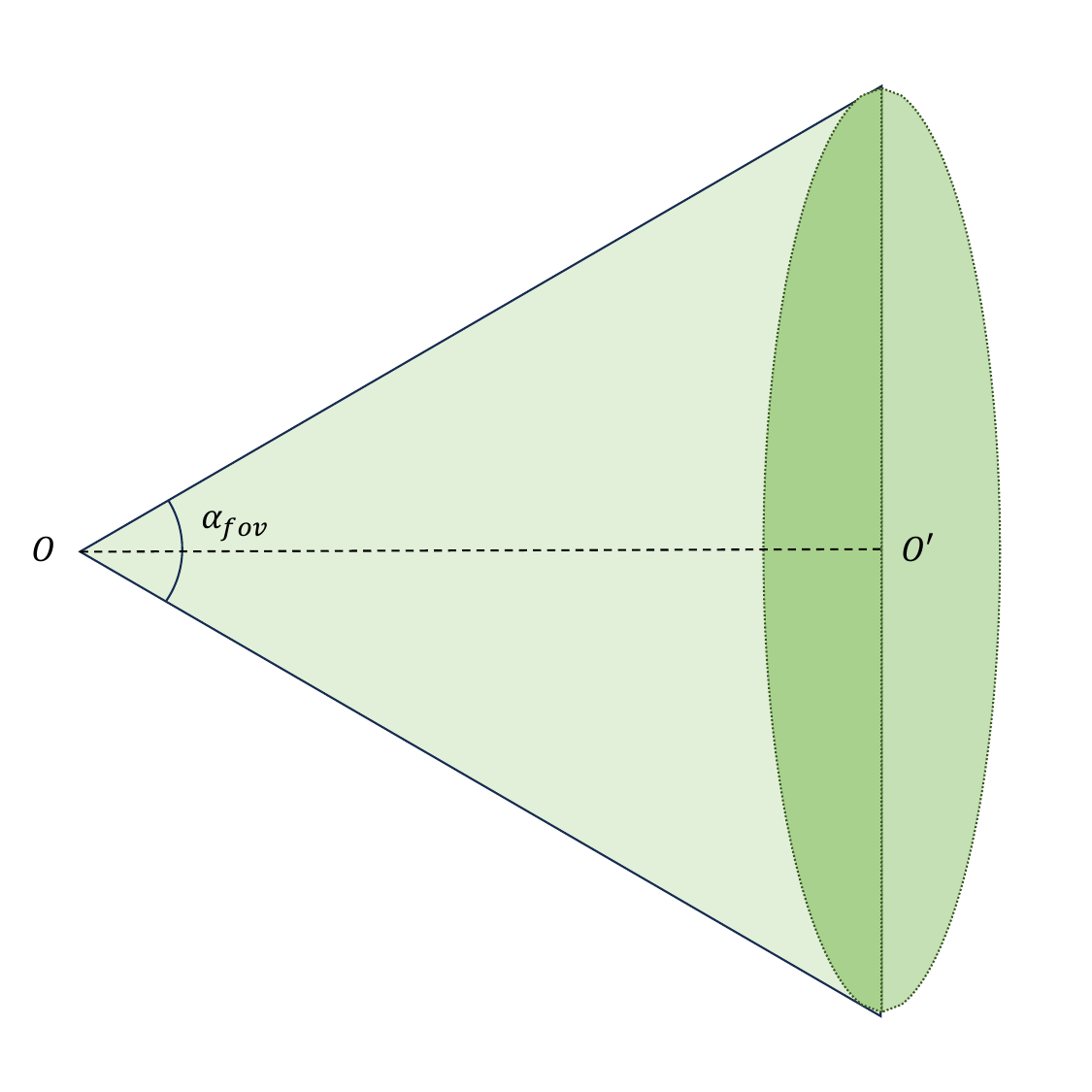}
\caption{ The field of view. }
\label{fig10}
\end{figure}

Next, let us discuss the field of view. 
It is characterized by the angles in the planes $ yO'z $ and $ xO'z $. 
For convenience, we consider these two angles are equal.
An example of the field of view is shown in Fig. \ref{fig10}. 
To determine the dimensions of the square screen, we need to calculate its length. 
The length $ L $ can be expressed as:
\begin{align}
L=2|\overrightarrow{OP}|\tan\frac{\alpha_\text{~fov}}{2}.
\end{align}

We consider $ n\times n $  pixels on the screen, where the length occupied per pixel is given by
\begin{align}
\ell=\frac{2|\overrightarrow{OP}|}{n}\tan\frac{\alpha_\text{~fov}}{2}.
\end{align}
The pixels are labeled by $ (i,j) $, with the pixel at the bottom left corner being $ (1,1) $ and the pixel at the top right corner being $ (n,n) $.
The indices $ i $ and $ j $ range from $ 1 $ to $ n $.
For the Cartesian coordinates of the center point of a pixel, we have
\begin{align}\label{B7}
x_{P'}=\ell \left(i-\frac{n+1}{2}\right), && y_{P'}=\ell\left(j-\frac{n+1}{2}\right).
\end{align}
By comparing equations \meq{B4} and \meq{B7}, we can establish a relationship between $ (i,j) $ and $ (\theta,\Psi) $ as follows:
\begin{align}
\tan\Psi&=\frac{j-(n+1)/2}{i-(n+1)/2},\\
\tan\frac{\theta}{2}&=\tan\frac{\alpha_\text{~fov}}{2}\frac{\sqrt{[i-(n+1)/2]^2+[j-(n+1)/2]^2}}{n}.
\end{align}

\begin{figure}[H]
\centering
\includegraphics[width=0.9\linewidth]{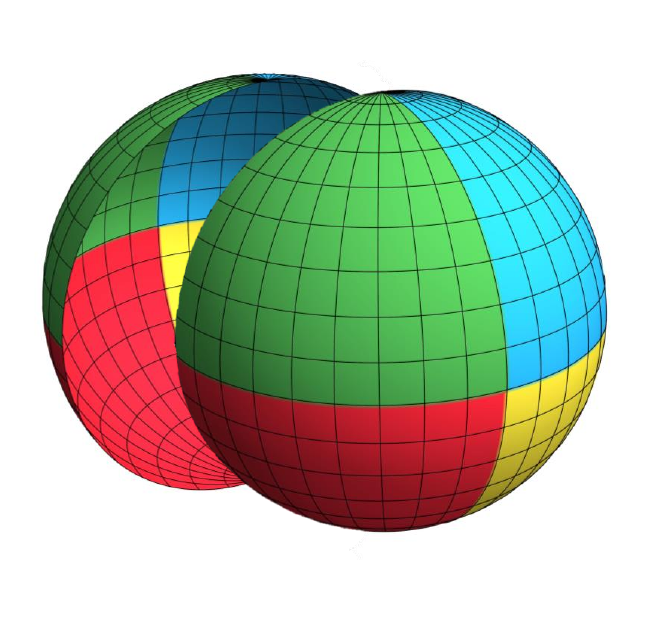}
\caption{Illustration of our spherical light source at infinity.}
\label{fig11}
\end{figure}
Finally, we state that we use an extended source to illuminate the system.
The model, illustrated in Fig. \ref{fig11}, displays a cutaway view of the photosphere, exposing its internal structure.
We segment the sphere into a grid network formed by lines of latitude and longitude, with an interval of $ \pi/18 $ between each pair of adjacent lines. 
To render an image of the black hole, we assign a specific color to each grid segment. 
The extended light source is divided into four distinct colors. 
The color of each pixel on the image is determined by tracing the path of photons from the corresponding points on the extended source. 
Areas that appear dark signify that the photons have been absorbed by the black hole.
\\~~

%\bibliography{Myreftitle}

\end{document}